\documentclass[conference,compsoc]{IEEEtran}
\usepackage{graphicx}
\usepackage{enumitem} 
\usepackage{amsmath}
\usepackage{amssymb}
\usepackage{array}
\usepackage{url}

\usepackage{algorithmic}

\ifCLASSOPTIONcompsoc
  \usepackage[nocompress]{cite}
\else
  \usepackage{cite}
\fi

\ifCLASSOPTIONcompsoc
 \usepackage[caption=false,font=footnotesize,labelfont=sf,textfont=sf]{subfig}
\else
 \usepackage[caption=false,font=footnotesize]{subfig}
\fi

\ifCLASSINFOpdf
\else
\fi

\usepackage{fixltx2e}
\usepackage{stfloats}
\usepackage{url}

\begin{document}

\title{Robust and Precise Application Fingerprinting on 5G Physical Uplink Channel}

\author{Yu Li, Liqi Zhuang, Dong Wei, Jiwen Luo, Hang Zhang, Meng Zhang, Xiaona Li, and Weiqing Huang}

\maketitle

\newcommand{\name}{\textit{Crosshair}}

\begin{abstract}

Air fingerprinting infers application activity by sniffing metadata from cellular control channels. 5G encrypts these channels, breaking the attack chain that prior attacks depend on. This paper reveals a novel physical-layer side channel that bypasses encryption: under the link adaptation mandated by the cellular communication standard, the uplink Modulation and Coding Scheme (MCS) remains stable, so the number of Physical Resource Blocks (PRBs) occupied by a transmission accurately reflects the IP packet length. Combined with the uplink control channel that carries downlink information, an attacker can reconstruct a bidirectional traffic profile.  This bidirectional information recovery can be achieved simply by observing the uplink spectrum, without decoding any channel. Building on this side channel, we design \name{}, a passive three-step attack. First, a blind extraction stage recovers the uplink physical channel occupancy from raw IQ samples via energy detection, reconstructing bidirectional traffic from uplink spectrum. Second, we design a data augmentation method that synthesizes spectral profiles across diverse channel conditions, eliminating the need for prior knowledge of the communication environment. Third, cross-modal alignment embeds the spectral and IP domains into a shared space, enabling new applications to be enrolled from a collected IP trace alone. Extensive experiments on a 5G New Radio (NR) testbed demonstrate the robustness and precision of \name{}: it outperforms the State-of-the-Art (SOTA) physical layer fingerprinting method in application recognition accuracy, and maintains high accuracy in cross-MCS scenarios.

\end{abstract}

\IEEEpeerreviewmaketitle

\section{Introduction}\label{sec:1}

Cellular networks transmit metadata (e.g., packet timing, size, and direction) on physical control channels. On 4G LTE, this metadata travels unencrypted over the air interface, enabling attackers to infer application usage by parsing unencrypted control-plane messages. A mature attack chain of 4G air fingerprinting has been established, including identity tracking~\cite{8835335, 11304648, hong2026passive}, traffic sniffing~\cite{hoang:ltesniffer, 236354, 10.1145/3495243.3560525}, and privacy exposure through app fingerprinting~\cite{Balasingam2017PosterBL, 10202613, 10.1007/978-981-96-9872-1_4, 9631415}, website fingerprinting~\cite{11145, 8835335}, video fingerprinting~\cite{bae:2022:watching}, service fingerprinting~\cite{10181886, 9210554, 11047734}, and device fingerprinting~\cite{9003304, 9500470}. However, these attacks all rely on the same prerequisites that 4G control channels are unencrypted, exposing metadata to any adversary equipped with a radio analyzer.

Extending this air attack model to 5G NR faces three obstacles. First, \textbf{identity concealment}: the identity (Subscription Permanent Identifier, SUPI) of the 5G user is encrypted. Although an attacker can capture encrypted SUPI~\cite{10.1145/3448300.3467826}, it cannot be linked to the user's temporary identity (Radio Network Temporary Identifier, RNTI) used over the air interface, nor can it be resolved to the subscriber's real identity (International Mobile Subscriber Identity, IMSI). Second, \textbf{downlink unobservability}: beamforming concentrates downlink transmission into a narrow spatial beam. A deviation of only a few degrees renders the downlink signal unusable for reliable sniffing~\cite{8371237, 9034044}. Third, \textbf{channel encryption}: the control channel for metadata acquisition is scrambled with RNTI, which is only exposed during the random access procedure~\cite{10.1145/3680121.3697808, 10.5555/3766078.3766355}. When using mobile phones, this procedure will not be triggered unless the cellular network is reset. Several works have attempted to port air fingerprinting to 5G NR~\cite{11034666, 10723461, 11304648, 10.1145/3448300.3467826, DBLP:conf/ndss/HussainECLB19, DBLP:journals/corr/abs-2512-20622}, but they either require privileged access to base station (BS) or user equipment (UE, e.g., mobile phone)~\cite{10723461, 10757717}, or claim 5G applicability solely by extrapolation from 4G without empirical validation.

Our insight is that shifting perspective to uplink fundamentally changes the attacker's position. UE's antenna radiates omnidirectionally, making its uplink signal observable from any angle regardless of BS's beamforming. attacker remains entirely passive, collecting raw In-phase and Quadrature (IQ) samples without actively transmitting, interacting with the BS, or resolving any identity. We observe that even this purely passive observation carries enough information to recover traffic patterns. Within the uplink spectrum, we observe a physical layer side channel. The MCS that governs uplink transmission remains stable under link adaptation mandated by 3GPP TS~38.214. Although this mechanism is designed to achieve more efficient link adaptation, it can be exploited by attackers for privacy inference.

To effectively utilize the discovered side channel, we recall three challenges that have emerged from previous air fingerprinting attacks. First, \textbf{insufficient precision:} the closest prior work~\cite{11034666} infers traffic from ACK/NACK messages on the Physical Uplink Control Channel (PUCCH). ACK/NACK is inherently a binary signal that indicates whether a downlink Transport Block (TB) was decoded, but carries no information about how much data was transmitted. Furthermore, 5G allows PUCCH to be multiplexed onto Physical Uplink Shared Channel (PUSCH) under high load conditions, which can make metadata inaccessible to a spectrum observer. Second, \textbf{environment dependence:} previous work collects and evaluates datasets under a single, fixed channel condition. In practice, an attacker typically lacks prior knowledge of specific channel conditions of the target environment (e.g., a private bedroom or an office). Obtaining such knowledge would require physically accessing these locations and deploying sniffers to characterize the channel in advance, which elevates the threat model to an unrealistically powerful adversary. Third, \textbf{prohibitive data collection cost:} a practical obstacle to deploying air fingerprinting attacks is the cost of spectrum capture. Unlike IP traffic collection, which can be performed by running a script on a commodity computer~\cite{SeleniumHQ2026}, spectral capture generates massive data volumes. For instance, at a 30.72\,MHz sampling rate, every 30 seconds of capture produces nearly 1\,GB of data. This makes dataset maintenance and updates prohibitively expensive.

In this paper, we design \name{}, a passive fingerprinting attack that utilizes a novel side channel to identify applications. \name{} follows a three-stage pipeline. First, a blind spectral demultiplexing stage extracts PUSCH bandwidth and PUCCH activity from raw IQ samples via energy detection, forming a more precise bidirectional spectral traffic profile without decoding any messages (Section~\ref{sec:method-extract}). Second, we design specialized data augmentation for unknown communication environments and deterministically translate air traffic traces collected in the laboratory to span different channels, thereby replacing prior knowledge of channel conditions (Section~\ref{sec:method-twin}). Third, a cross-modal alignment stage learns a shared embedding space between IP and spectral domains, so that enrolling a new application requires only IP trace collection rather than repetitive Radio-Frequency (RF) collection (Section~\ref{sec:method-learning}). We evaluate \name{} on a 5G NR testbed with commercial UEs, demonstrating robust and precise application recognition at over 90\% accuracy across diverse channel conditions.

The main contributions of this paper are:
\begin{itemize}
\item We identify a novel and unnoticeable physical-layer side-channel in 5G NR. According to our best knowledge, \name{} is the first 5G air fingerprint attack to utilize this side channel.
\item We design a blind spectral demultiplexing pipeline that extracts PUSCH bandwidth and PUCCH activity from raw IQ samples, reconstructing precise bidirectional traffic from unidirectional uplink spectrum capture.
\item We design a data augmentation for cross-environment evaluation, enabling scalable training data generation without any prior knowledge of the target's channel conditions.
\item We achieve the first cross-modal zero-shot fingerprint in the 5G physical layer. We develop an alignment framework to register a new application without re-collection of the air dataset.
\item We implement and evaluate \name{} on a 5G NR testbed with commercial UEs, demonstrating application recognition accuracy exceeding 90\% across diverse channel conditions.
\end{itemize}

\section{Background}\label{sec:mcs}
\subsection{Spectrum Resources}\label{sec:channel_intro}
The 5G NR air interface is organized into physical channels. BS communicates with UE through these channels, each serving a distinct function in the communication protocol.

\begin{figure}[htbp]
  \centering
  \includegraphics[width=0.7\linewidth]{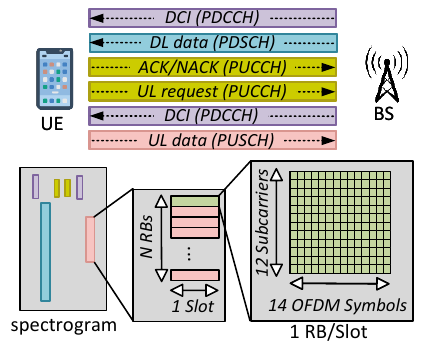}
  \caption{Illustration of 5G NR air interface spectrum resources.}\label{fig:background}
\end{figure}

\begin{itemize}
\item \textbf{Physical Broadcast Channel (PBCH): } BS broadcasts the Master Information Block (MIB) on this channel, containing basic cell configuration. The Primary Synchronization Signal (PSS) and Secondary Synchronization Signal (SSS) are also transmitted here. UEs use these unencrypted signals for cell search, time-frequency synchronization, and initial access.

\item \textbf{Physical Downlink Control Channel (PDCCH): } This channel carries Downlink Control Information (DCI), which specifies scheduling grants (MCS, PRB allocation, slot assignment) for downlink and uplink data transmissions. In a 5G cellular network, DCI is scrambled with the target UE's RNTI, making it unreadable to passive eavesdroppers.

\item \textbf{Physical Downlink Shared Channel (PDSCH): } This channel carries downlink user data, System Information Blocks (SIBs), and paging messages. PDSCH transmissions are scheduled by the associated PDCCH.

\item \textbf{Physical Random Access Channel (RACH): } UE uses this channel to initiate the random access procedure when connecting to the network. Upon successful access, the BS assigns an RNTI used to scramble subsequent DCI. Therefore, DCI can only be decoded by a sniffer that observes the random access exchange.

\item \textbf{PUCCH: } This channel carries Hybrid Automatic Repeat Request (HARQ) ACK/NACK feedback for downlink transmissions. Crucially, the density of PUCCH activity reflects the intensity of downlink traffic, which is a property that the attacker exploits to reconstruct the downlink traffic pattern without any decoding.

\item \textbf{PUSCH: } This channel carries uplink user data and may also carry Uplink Control Information (UCI) when multiplexed with data. PUSCH transmissions are scheduled by PDCCH grants. The bandwidth occupied by a PUSCH transmission is directly determined by MCS and the number of allocated PRBs.

\end{itemize}

Figure~\ref{fig:background} illustrates the time-frequency resource grid. PDCCH grants (purple) indicate slot, MCS, and PRB allocation for each subsequent data transmission. Each uplink transmission (pink) occupies one slot, whose duration depends on Subcarrier Spacing (SCS) configured by the BS (Table~\ref{tab:scs-slot}). Each PRB (green) spans 12 subcarriers in the frequency domain. The total bandwidth occupied by $N$ PRBs is therefore $12 \cdot {SCS} \cdot N$\,kHz. The data within these PRBs is modulated according to MCS and transmitted on orthogonal subcarriers via Orthogonal Frequency Division Multiplexing (OFDM).

\begin{table}[htbp]
\centering
\small
\caption{5G NR SCS Configurations and Slot Duration}
\label{tab:scs-slot}
\begin{tabular}{cc}
\hline
\textbf{SCS (kHz)} & \textbf{Slot Duration (ms)} \\
\hline
15  & 1      \\
30  & 0.5    \\
60  & 0.25   \\
120 & 0.125  \\
240 & 0.0625 \\
\hline
\end{tabular}
\end{table}

\subsection{Modulation and Coding Scheme}
\begin{figure}[!htbp]
  \centering
  \includegraphics[width=\linewidth]{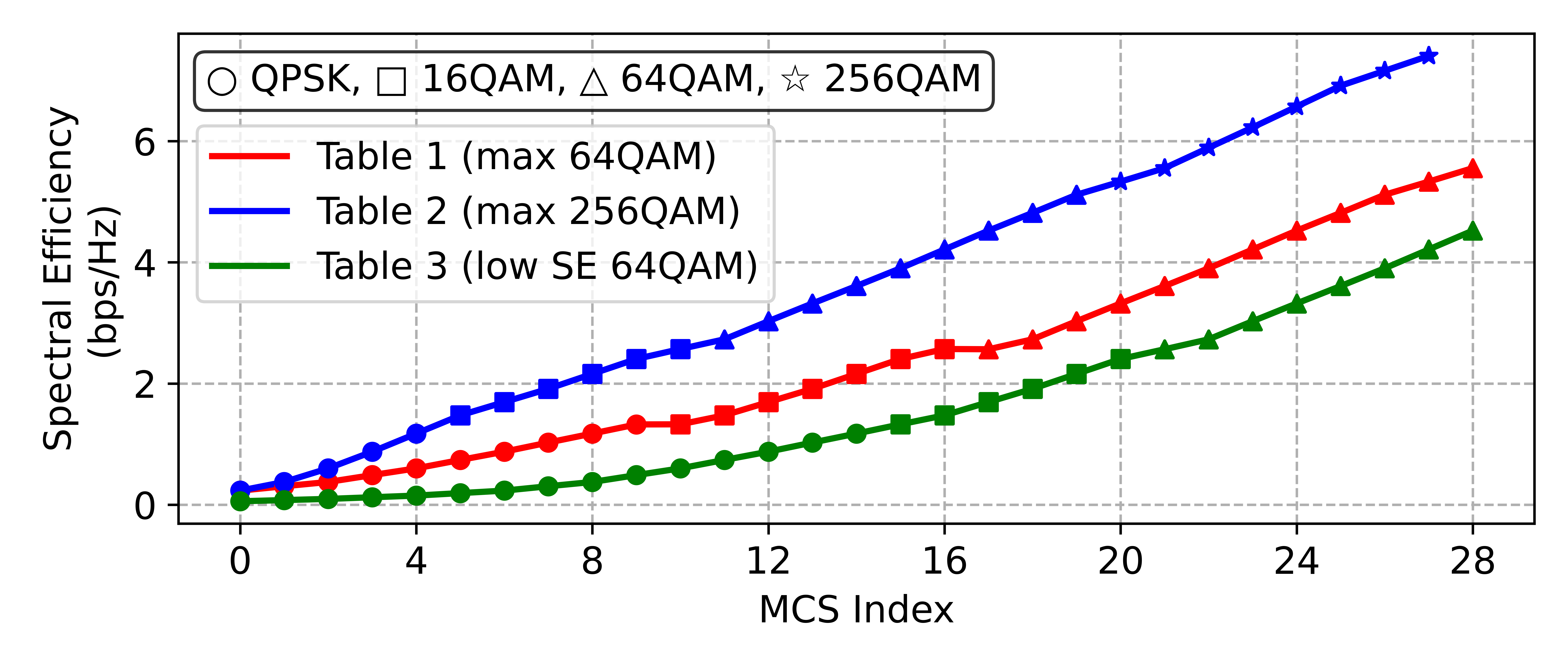}
  \caption{Variation of spectral efficiency with the MCS index.}\label{fig:mcs}
\end{figure}

Over the air interface, the fundamental transmission unit is TB, whose size (in bits) is jointly determined by the MCS index and the number of allocated PRBs, per mapping defined in 3GPP TS 38.214~\cite{3gpp.38.214}. MCS index is a 5-bit value that specifies modulation order $Q$ (QPSK, 16QAM, 64QAM, or 256QAM) and code rate $R$, yielding a spectral efficiency of $Q \times R$ bits/s/Hz. BS dynamically selects the MCS index to balance throughput against block error rate (BLER) based on the UE's reported channel quality. An appropriate MCS table is chosen from among three defined in TS 38.214 according to the UE's capability reported during RRC connection setup. Figure~\ref{fig:mcs} shows spectral efficiency as a function of MCS index for three tables. 256QAM table targets high-throughput scenarios, while the Low-SE 64QAM table is designed for Ultra-Reliable Low-Latency Communications (URLLC).

Thus, a passive observer sees two signals in the uplink spectrum: PUSCH bandwidth (uplink data volume) and PUCCH activity (downlink acknowledgment rate). The critical question is whether MCS is stable to ensure that the side channel of PRB count directly reflects TB size (TBS).

\section{Motivation}\label{sec:motiv}
On 4G LTE, attackers can trivially correlate air signals with user traffic: unencrypted DCI exposes the TBS of each transmission, directly revealing the packet length pattern. LTESniffer~\cite{hoang:ltesniffer} and similar tools operationalize this approach. 5G NR encrypts DCI so that it is impractical for a passive adversary to obtain TBSs from downlink control plane messages.

However, TBS is not the only quantity that leaves a footprint on the spectrum. Under nearly stable and continuous MCS, the number of PRBs allocated to a transmission is proportional to its TBS (Section~\ref{sec:mcs}), and PRB count directly determines occupied bandwidth. A passive adversary without decoding DCI can still observe \textit{how many} PRBs are active in each slot by inspecting energy distribution in the spectrogram. The critical question is whether MCS remains stable enough for PRB count to serve as a reliable proxy for TBS. This section investigates this question through empirical measurement, numerical analysis, and protocol-level reasoning.

\textbf{Empirical observation.} The first question is whether uplink MCS remains stable under real-world conditions, without which PRB count cannot reliably proxy TBS. To answer this, we conducted controlled measurements in which we temporarily instrumented a test UE to log per-slot MCS assignments. In particular, we allowed participants to freely operate their phones to collect real-world data, including natural actions such as rotating the device, hand manipulation, and walking freely indoors, depending on each participant's usage habits. We do not consider rapid mobility scenarios such as running, cycling, or driving while using the phone, as the sniffer would lose the signal once the UE moves beyond detection range. We tested 4 UEs (iPhone~12, Xiaomi~12, Samsung Galaxy~S23, OPPO Find~X7) running different services on two outdoor commercial BSs. Measurement required briefly placing the UE in airplane mode to trigger a random access event that exposes RNTI for DCI decoding; this procedure is feasible only under controlled conditions and is not part of \name{}'s attack methodology.

\begin{figure*}[htbp]
    \centering
    \subfloat[Brand:iPhone, type: video call, BS: 1, period: morning.]{\includegraphics[width=0.49\linewidth]{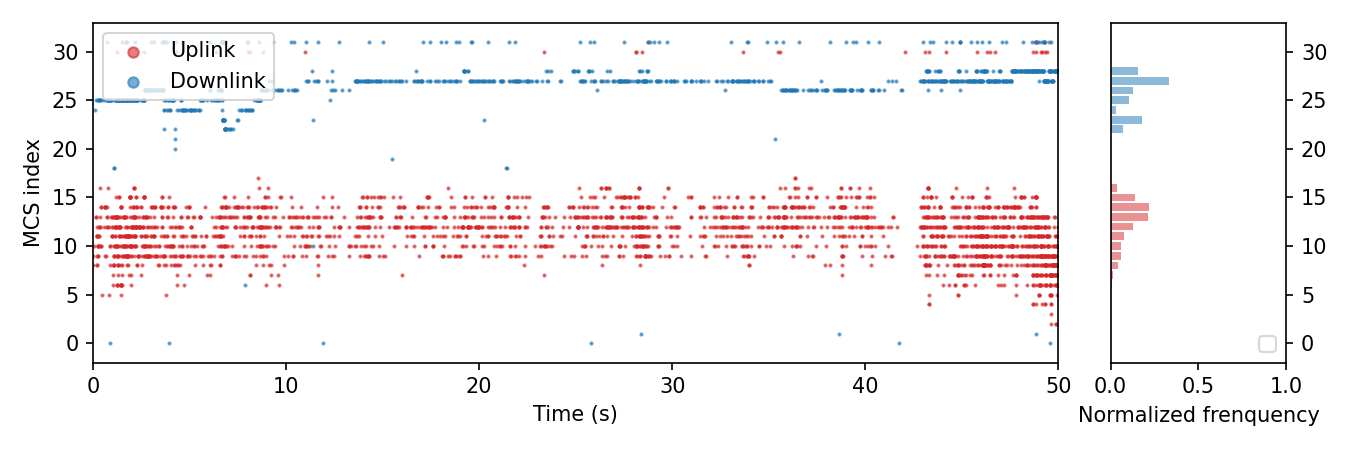}}
    \hfill
    \subfloat[Brand:OPPO, type: live stream, BS: 1, period: morning.]{\includegraphics[width=0.49\linewidth]{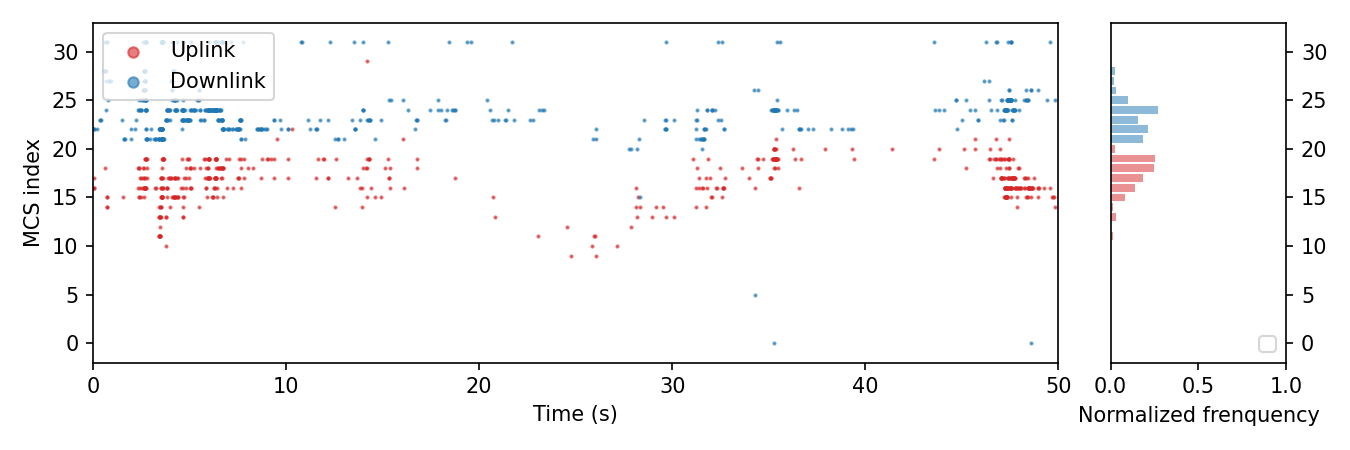}}
    
    \subfloat[Brand:Samsung, type: live stream, BS: 2, period: middle.]{\includegraphics[width=0.49\linewidth]{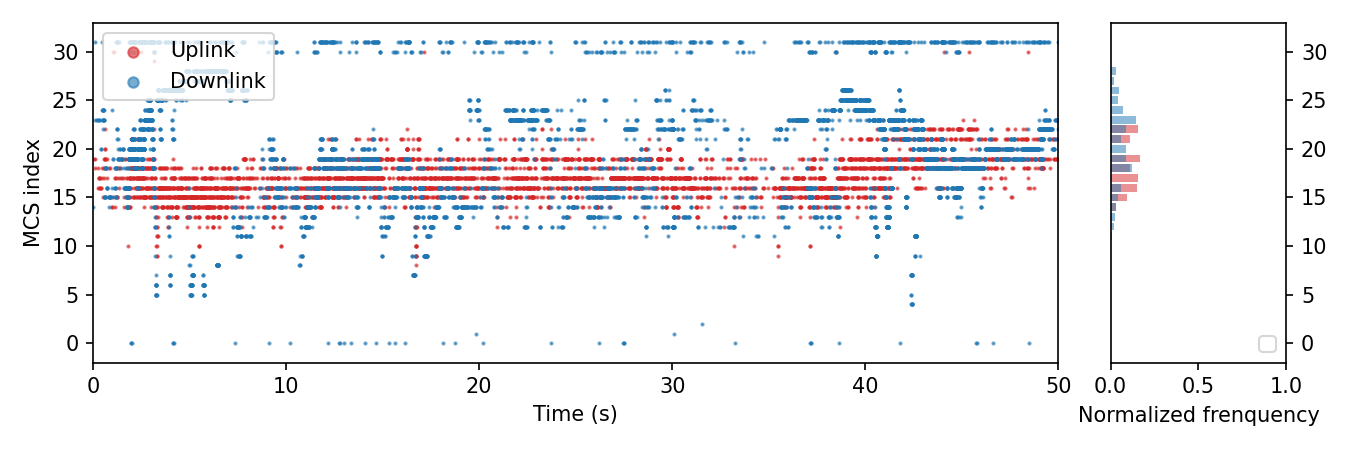}}
    \hfill
    \subfloat[Brand:Xiaomi, type: video conference, BS: 1, period: evening.]{\includegraphics[width=0.49\linewidth]{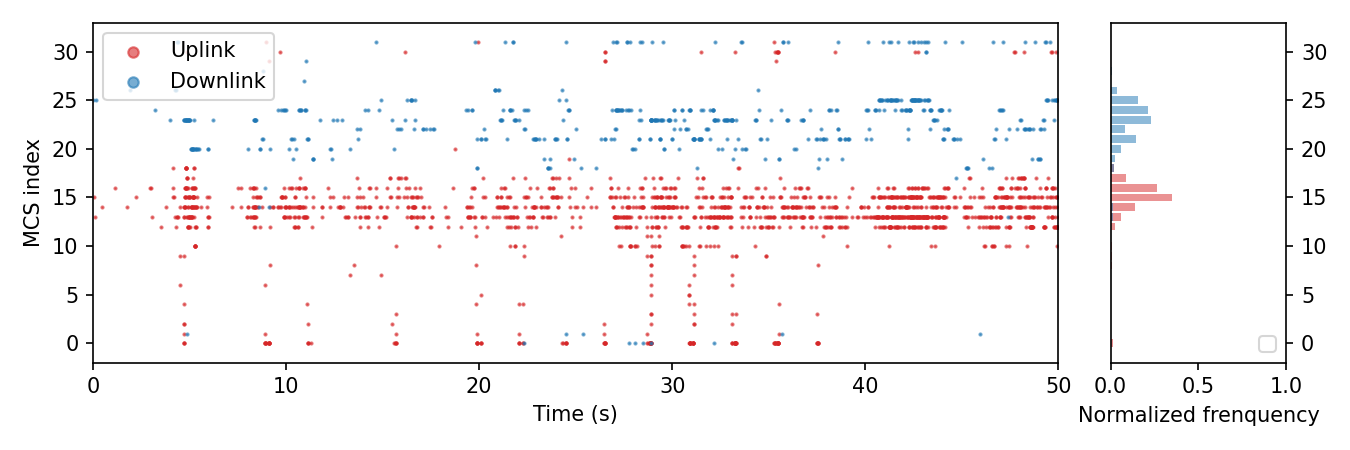}}
    
    \caption{Curve of MCS index over time and histogram of MCS index under different UEs at real BSs.}
    \label{fig:ues} 
\end{figure*}

Figure~\ref{fig:ues} shows MCS time series and corresponding histograms across four UEs, two traffic types, and two commercial BSs. We observe that uplink MCS remains concentrated within a narrow band in all conditions. Over 90\% of MCS values fall within 5 adjacent indices, regardless of UE brand, application, time of day, or BS. In particular, MCS concentration persists across the majority of typical phone usage scenarios, including walking. Furthermore, uplink MCS is consistently lower than downlink MCS, meaning that uplink PRB fluctuations have proportionally less impact on data rate. We also find that stability persists across both heavy (video conferencing) and light (live streaming) uplink traffic, indicating that the effect is driven by channel conditions and scheduler policy rather than traffic volume.

\textbf{Numerical analysis.} MCS stability limits the MCS index to a small discrete set, but does not eliminate variation entirely. We quantify residual distortion through \textit{maximum spectral contraction ratio}, which is defined as the largest bandwidth ratio observed when transmitting an identical TBS under two different MCS indices within the stable range. Let $S = \max(\text{MCS}) - \min(\text{MCS})$ denote MCS fluctuation. For a given TBS and MCS index, the required PRB count follows the TBS-to-PRB mapping in TS~38.214.

\begin{figure}[htbp]
    \centering
    \subfloat[Table 1 (max 64QAM).]{\includegraphics[width=0.495\linewidth]{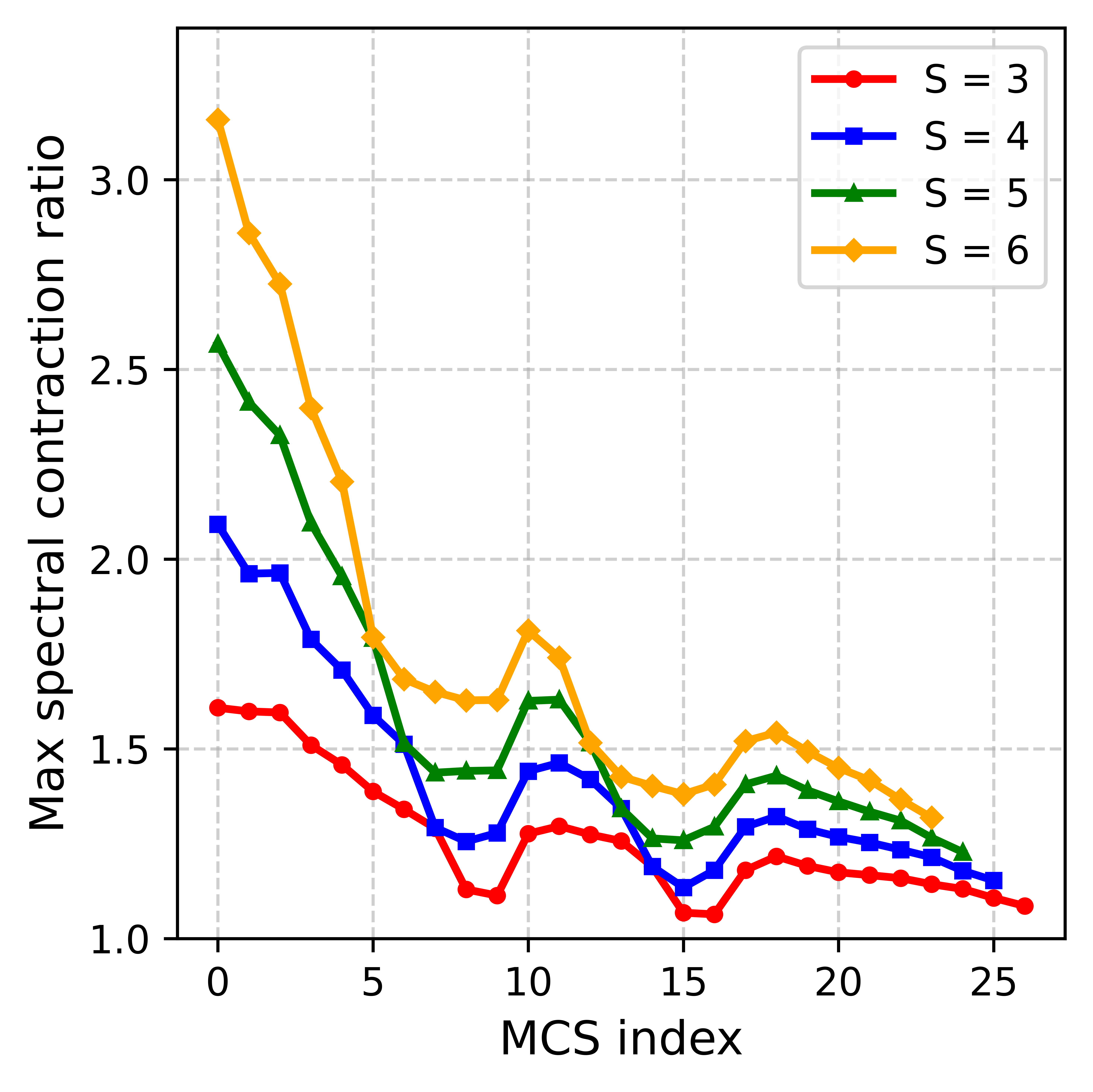}}
    \hfill
    \subfloat[Table 2 (max 256QAM).]{\includegraphics[width=0.495\linewidth]{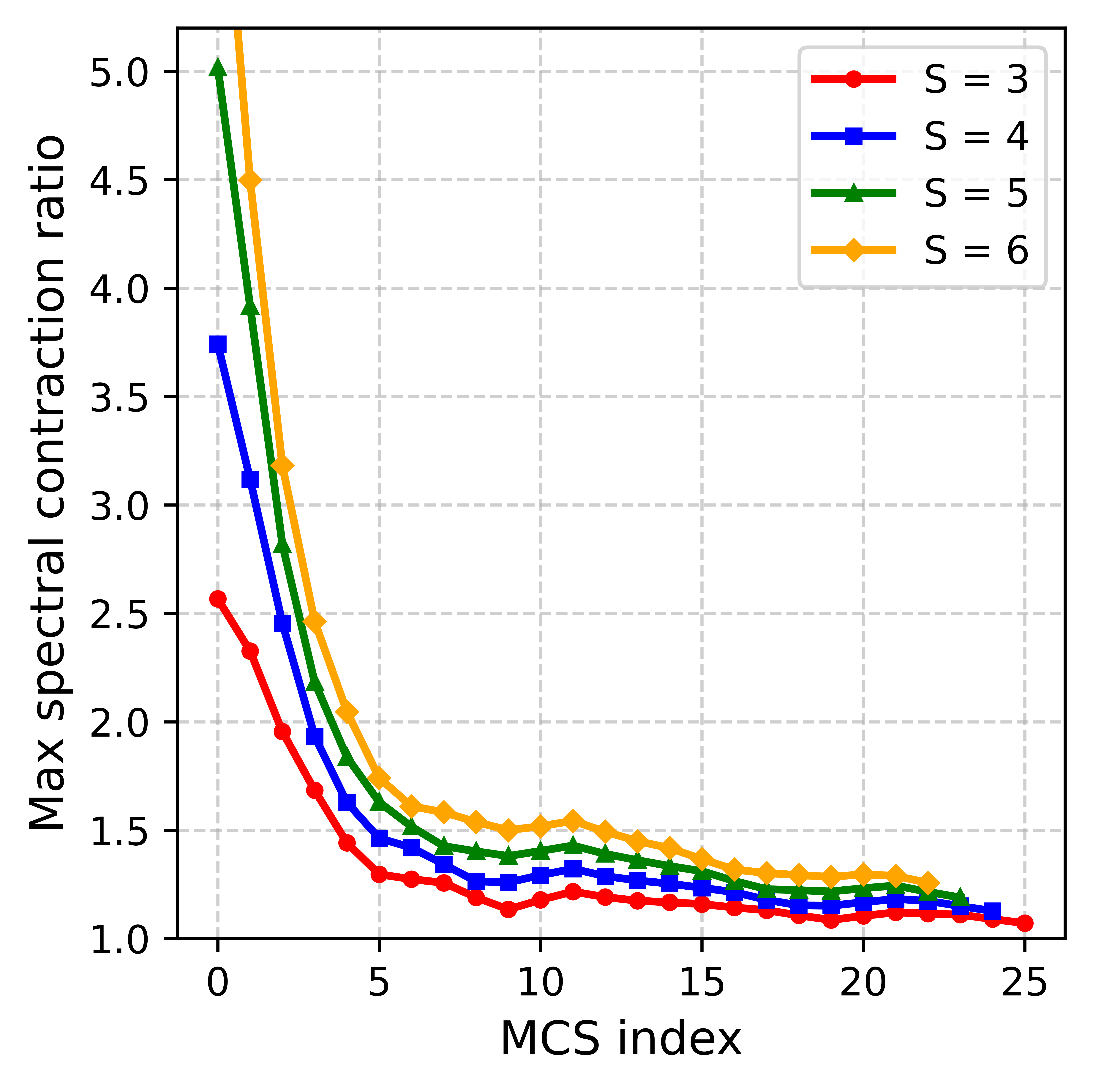}}
    \caption{Maximum spectral contraction ratio versus MCS index for two 5G MCS tables under different MCS fluctuation levels $S$.}
    \label{fig:shrink}
\end{figure}

Figure~\ref{fig:shrink} shows the spectral contraction ratio across the two most commonly deployed 5G MCS tables (64QAM and 256QAM) under increasing $S$. As the minimum MCS index increases, the contraction ratio decreases. Even under substantial fluctuation ($S=6$), provided the minimum MCS exceeds 5, the spectral contraction ratio remains below $2\times$. This means that PRB count varies by at most a factor of two when MCS fluctuates within its empirically observed range. Moreover, large TBs are typically transmitted using higher MCS indices, so the effect is even smaller at higher MCS indices, where spectral efficiency saturates. Consequently, PRB occupancy sequence $W[t]$ preserves the temporal morphology of the underlying TB sequence, distorted only by a bounded scaling factor. This bounded distortion is a property that enables blind traffic inference from a spectrogram.

\begin{figure}[htbp]
  \centering
  \includegraphics[width=\linewidth]{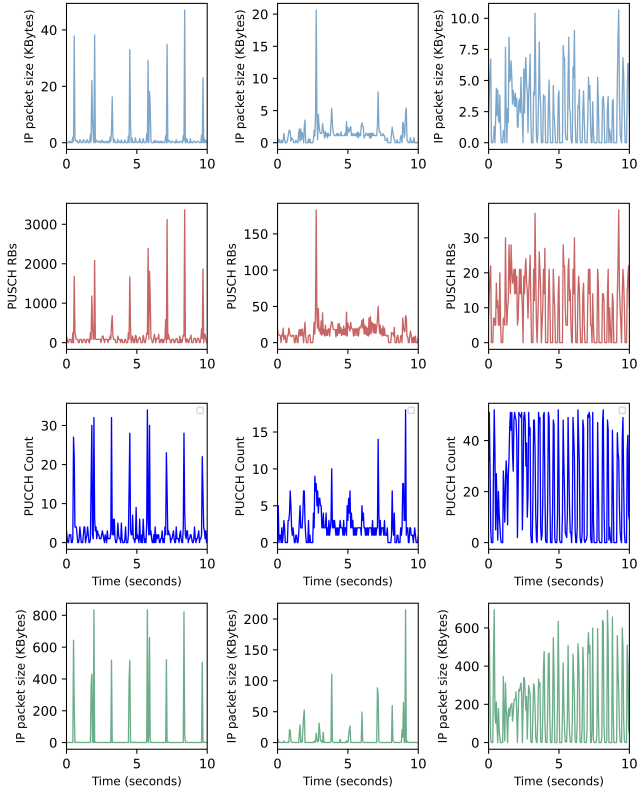}
  \caption{Comparison of characteristics of 5G NR uplink physical signals in different channel environments with original IP traffic. Left: Worst channel environment. Middle: Optimal channel environment. Right: Best channel environment.}\label{fig:ip_phy}
\end{figure}

Figure~\ref{fig:ip_phy} compares original uplink and downlink IP traffic traces against observations (PUSCH carrier width and PUCCH counts) from the spectrogram, across three channel conditions: worst channel (left), moderate channel (middle), and best channel (right). In all three regimes, PUSCH carrier width follows the uplink IP traffic envelope, and PUCCH activity density reflects the downlink traffic pattern. The bidirectional spectral traffic profile thus captures the essential structure of IP traffic, which is the side channel that \name{} exploits.

\begin{figure}[htbp]
  \centering
  \includegraphics[width=0.9\linewidth]{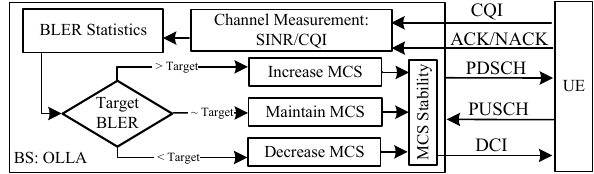}
  \caption{OLLA mechanism.}\label{fig:adjust}
\end{figure}

\textbf{Protocol analysis.} The observed stability is not a coincidence. It is a direct consequence of the 3GPP link adaptation mechanism \cite{3gpp.38.214, 3gpp.38.213}. The BS's outer-loop link adaptation (OLLA, Figure~\ref{fig:adjust}) maintains a target BLER (typically $0.1$) by applying a slowly varying Signal-to-Interference-plus-Noise Ratio (SINR) offset $\delta$ before MCS selection. On each successful decode, $\delta$ is decremented; on each failure, $\delta$ is incremented. The step-size ratio $\Delta_{\text{up}}/\Delta_{\text{down}} = (1-p_{\text{target}})/p_{\text{target}}$ ensures that $\delta$ converges to a stable equilibrium under stationary channel conditions. For the uplink, this effect is amplified by the UE's limited transmit power and hardware constraints, which restrict the SINR dynamic range and make the uplink MCS more conservative than the downlink.

Overall, based on our empirical observation, numerical analysis, and protocol-level reasoning, we summarize that the PRB occupancy sequence $W[t]$ extracted from the spectrogram is a reliable proxy for the IP traffic envelope. This is the side channel that \name{} exploits: by observing only the uplink spectrum, a passive adversary can reconstruct the bidirectional traffic pattern of the target user without decoding any message.

\section{Threat Model}

\begin{figure}[!htbp]
  \centering
  \includegraphics[width=0.7\linewidth]{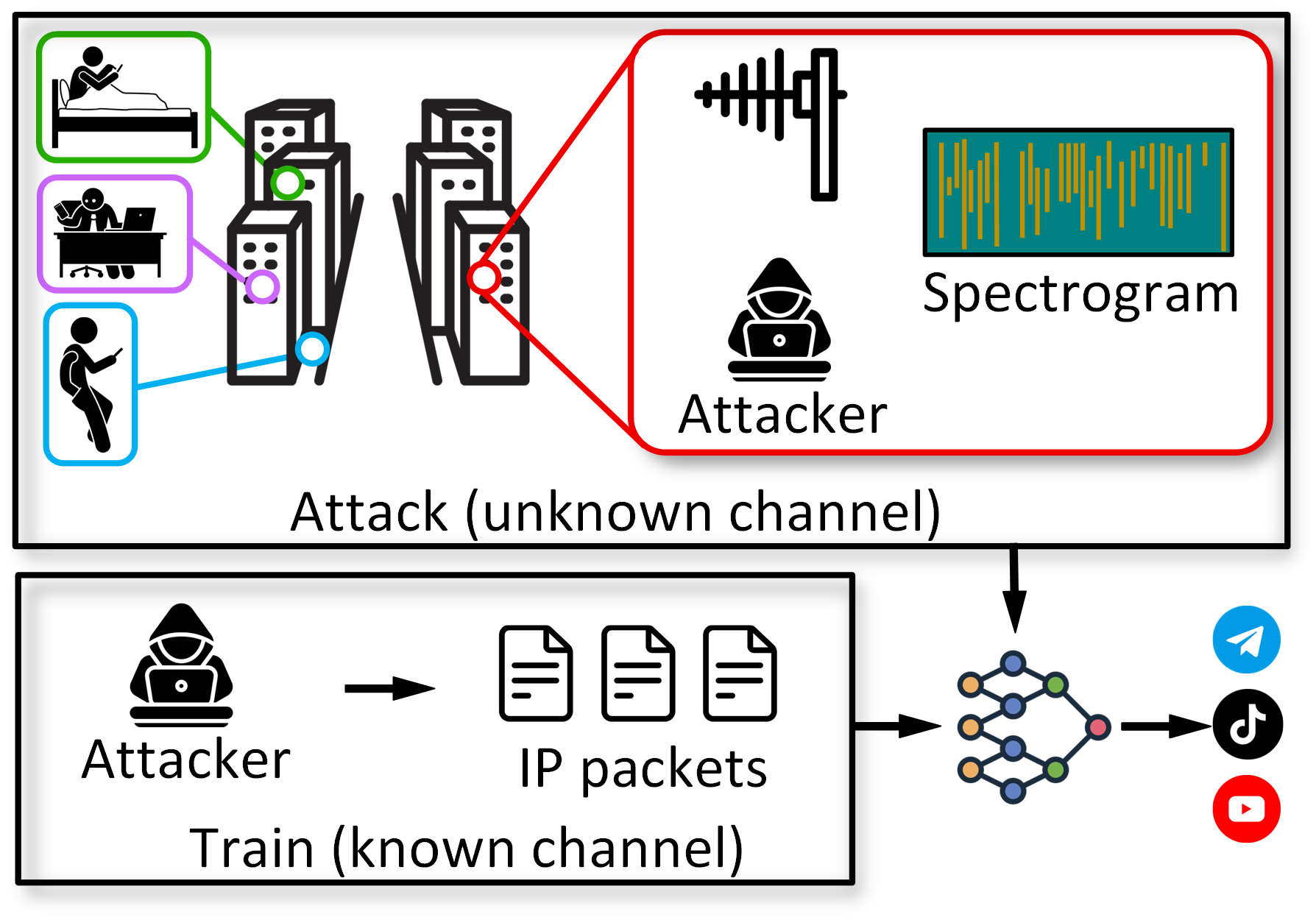}
  \caption{Threat model of \name{}.}\label{fig:threat}
\end{figure}

\noindent \textbf{Attack scenario.} We consider a passive third-party attacker who seeks to infer which application a target user is running by observing the user's uplink radio signals. A mobile phone user connects to a cellular network and uses online applications. The attacker deploys a receiver with a directional antenna near the victim, captures raw IQ samples of the uplink spectrum, and analyzes the resulting spectrogram (Figure~\ref{fig:threat}). The attacker is entirely passive: they cannot install malware on the UE, deploy a fake BS, parse any radio signal, collude with the MNO, or conduct active interventions such as challenge-response~\cite{10.1145/3448300.3467826} or communication-blocking attacks~\cite{10.5555/3766078.3766355}. In contrast to prior threat models, the attacker is \textbf{completely passive}.

\noindent \textbf{Attack capability.} During the train phase, the attacker can install a traffic capture application~\cite{PCAPdroid2026} on their own data collection phone or Android emulator to collect IP traces. At attack time, the attacker captures the uplink radio signal with a receiver placed near the victim. A low-cost Software Defined Radio (SDR), such as a Universal Software Radio Peripheral (USRP) B210, with a directional antenna is sufficient. The directional antenna suppresses co-channel interference from UEs at other angular positions, so the target's signal dominates the received spectrogram. The adversary must identify a location the victim frequents (e.g., an adjacent office or a nearby vehicle) and physically deploy the receiver there. No prior knowledge of the victim's identity, RNTI, or channel parameters is required.

\noindent \textbf{Attack assumption.} The attacker knows the carrier frequency, bandwidth, and duplexing mode of the target Mobile Network Operator (MNO), as these are publicly available and broadcast unencrypted in the MIB~\cite{DBLP:conf/ndss/ShaikSBAN16}. We assume the attacker can maintain reliable uplink reception for the duration of the monitoring period. We assume the user does not engage in high-speed movement beyond the sniffer's detection capability, as this would prevent the sniffer from acquiring the uplink radio signal. The method operates as long as the target's uplink signal is observable above the noise floor.



\section{Method}
In this section, we utilize the side channel observed in Section \ref{sec:motiv} to construct the pipeline of \name{}. 

\subsection{Overview of \name{}}
\begin{figure*}[!htbp]
  \centering
  \includegraphics[width=0.9\linewidth]{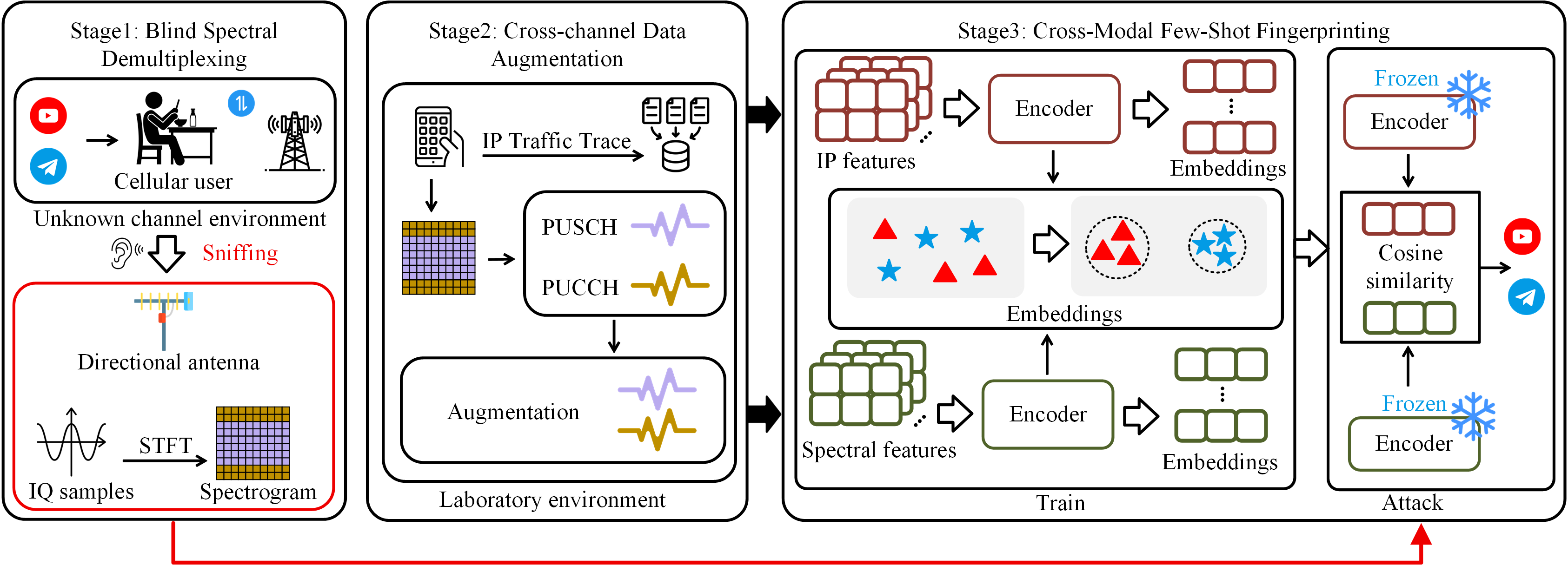}
  \caption{The overview of \name{}.}\label{eq:pucch}
\end{figure*}

In this paper, we design \name{}, a method that leverages the side channel to reconstruct bidirectional traffic from the uplink spectrum for precise and robust air fingerprinting. Specifically, \name{} captures the correlation between application IP traffic and uplink spectrum, enabling the inference of the application from a given spectrogram alone, even across unknown channel conditions. Figure \ref{eq:pucch} illustrates the overview of \name{}. \name{} consists of three modules designed to construct robust and precise air fingerprinting attacks, including blind spectral demultiplexing, cross-channel data augmentation, and cross-modal few-shot fingerprinting.


\textbf{Blind Spectral Demultiplexing} aims to extract and reconstruct bidirectional traffic from IQ samples captured by a radio sniffer, ensuring each application has a precise over-the-air feature representation. \name{} first converts raw IQ samples into a spectrogram via the Short-Time Fourier Transform (STFT). PUCCH and PUSCH can be separated since they occupy distinct regions in the spectrogram. Finally, \name{} leverages a directional antenna to isolate the target user's spectral footprint from interference and reconstruct bidirectional traffic from uplink channels (Section \ref{sec:method-extract}).

\textbf{Cross-channel Data Augmentation} aims to match the unknown communication environment using only data collected from the attacker's own environment, thereby reducing the cost of traffic collection and improving the robustness of feature extraction. \name{} models the MCS variation process as a mean-reverting random walk following the OLLA mechanism, generating synthetic MCS sequences under different channel conditions. It then computes the spectral contraction ratio and randomly scales each RB to the size corresponding to the target channel environment (Section~\ref{sec:method-twin}).

\textbf{Cross-modal Zero-shot Fingerprinting} learns the relationship between IP traffic and spectrogram sample pairs to align air traffic with IP traffic. An encoder encodes these spectral samples and compares them against a prototype library computed from prior IP traffic to infer the corresponding application. In this way, \name{} achieves scalable zero-shot attack without requiring retraining for new applications or large-scale prior air traffic collection (Section \ref{sec:method-learning}).

\subsection{Blind Spectral Demultiplexing}\label{sec:method-extract}

The attacker must extract the PRB time series from raw IQ samples. Prior 4G sniffer obtains exact PRB allocation by decoding PDCCH DCI. However, 5G DCI is scrambled with RNTI, which is unknown to a passive adversary and does not leak absent a random access event. 

Our insight is that decoding is unnecessary because aggregate PRB occupancy of the target user is directly observable through RB carrier width in the spectrogram, where each occupied PRB manifests as a localized energy concentration above the noise floor. Crucially, the attacker can further separate PUSCH and PUCCH by their frequency domain positions, as PUCCH resides at band edges while PUSCH occupies central contiguous allocation. Since PUSCH occupancy reflects uplink traffic volume and PUCCH activity reflects downlink acknowledgments, this procedure effectively reconstructs a bidirectional traffic profile from unidirectional spectrum capture. The entire pipeline isolates the target user's bidirectional traffic footprint from the spectrum signal using only energy detection and channel separation, without parsing any radio signals. We term this method \textbf{blind spectral demultiplexing}. The pipeline consists of three steps.

\textbf{Step 1: Spectrogram computation.} Given raw IQ samples at sampling rate $f_s$, we compute STFT. This yields a time-frequency energy map $E[t, f]$ where $t$ indexes OFDM symbols and $f$ indexes subcarriers.

We formalize this process as follows. The sniffer captures raw IQ samples as a complex-valued discrete signal
\begin{equation}
    x[n] = I[n] + j Q[n], \quad n = 0, 1, \ldots, N-1
\end{equation}
where $I[n]$ and $Q[n]$ denote in-phase and quadrature components, $f_s$ is sampling rate, and $N$ is total number of samples. STFT converts these IQ samples into a time-frequency representation by applying a sliding window aligned to the OFDM symbol boundary
\begin{equation}
    X[t, f] = \sum_{n=0}^{N_{\text{FFT}}-1} x[n + t \cdot H] \, w[n] \, e^{-j 2\pi f n / N_{\text{FFT}}}
\end{equation}
where $w[n]$ is a window function, $N_{\text{FFT}}$ is the Fast Fourier Transform (FFT) length configured to match the OFDM symbol duration determined by SCS, and $H$ is the hop length. The spectrogram is obtained from the squared magnitude
\begin{equation}
    E[t, f] = |X[t, f]|^2
\end{equation}
yielding a two-dimensional matrix in which $t$ indexes time frames and $f$ indexes frequency bins, with each element $E[t, f]$ representing the energy intensity at time $t$ and frequency $f$.

\textbf{Step 2: Bidirectional traffic reconstruction.} We now show how bidirectional traffic is reconstructed from uplink spectrum alone. Since PUCCH is itself an uplink physical channel, it is naturally observable alongside PUSCH in the same spectrum capture. By jointly monitoring PUSCH and PUCCH, we reconstruct bidirectional traffic from uplink spectrum alone, without ever decoding downlink.

Distinguishing these two channels in the spectrogram, however, is non-trivial. In 4G LTE, PUCCH is strictly allocated at the band edges flanking PUSCH, making the two channels straightforward to distinguish. In 5G NR, PUSCH and PUCCH are flexibly scheduled across the spectrum. However, from our empirical measurements on commercial 5G NR BSs, we observe that PUCCH also consistently resides at several fixed positions near the band edges, and does not fully comply with the protocol specification. We conjecture that BS designers adopt this simplified pattern for scheduling simplicity and cost considerations, aiming to maximize spectrum utilization. Therefore, we adopt a simple classification rule whereby slots with a single active RB at the band edges are labeled as PUCCH (marked as down) and slots with more than one active RB at the band center are labeled as PUSCH (marked as up). For each channel type $c \in \{\text{up}, \text{down}\}$, let $\mathcal{F}_c$ denote the set of RB indices belonging to that region. PRBs whose per-slot mean energy exceeds a noise-floor threshold are classified as active:

\begin{equation}
    \mathcal{A}_c[t] = \{ k \in \mathcal{F}_c \;|\; \bar{E}[t, k] > \tau \}, \quad c \in \{\text{up}, \text{down}\}
\end{equation}
where $\bar{E}[t, k]$ is the mean energy in RB $k$ during slot $t$, and $\tau$ is estimated from silent periods. This energy detection yields two independent active PRB sets without inspecting any DCI.

\textbf{Step 3: Target separation.} When multiple UEs transmit concurrently, the scheduler allocates non-overlapping PRB sets to each UE. The spectrogram therefore contains multiple distinct energy clusters, but without decoding DCIs the attacker cannot determine which cluster belongs to the target user. \name{} resolves this attribution problem via spatial filtering: the attacker orients a directional antenna toward the target user's known location. Since the target UE lies within the main lobe while co-channel interferers are attenuated by the directional gain roll-off, the target's PRB cluster dominates the received energy. The attacker then rejects background RF noise to harvest the target signal without requiring RNTI-based identity resolution. The result is a clean PRB occupancy sequence for the target user, extracted solely from energy measurements and spatial filtering.

Finally, aggregating the active PRBs per channel yields a pair of univariate time series:
\begin{equation}
    W_c[t] = |\mathcal{A}_c[t]|, \quad c \in \{\text{up}, \text{down}\}
\end{equation}
$W_{\text{up}}[t]$ captures the per-slot PUSCH occupancy, a direct proxy for the uplink packet length sequence, since $W_{\text{up}}[t] \propto \text{TBS}[t]$ under stable MCS (§\ref{sec:motiv}). $W_{\text{down}}[t]$ captures the per-slot PUCCH resource activity. Together, $(W_{\text{up}}[t], W_{\text{down}}[t])$ forms a \textbf{bidirectional spectral traffic profile}  reconstructed entirely from uplink spectrum capture. The entire pipeline requires no parsing. We denote the combined profile as $W[t] = (W_{\text{up}}[t], W_{\text{down}}[t])$.

\subsection{Cross-channel Data Augmentation}\label{sec:method-twin}

We now consider a practical obstacle from the attacker's perspective. The attacker does not know the target user's channel environment a priori, yet training data must be collected under comparable channel conditions to be effective. Exhaustively sweeping across all possible channel regimes is prohibitively laborious, and certain channel conditions may simply not arise during any feasible collection window. Therefore, we address this issue through data augmentation. Data augmentation has been widely adopted in image recognition~\cite{11094472, DBLP:conf/aaai/LuXWDH26} and related domains to improve model generalization and robustness. In our scenario, we design an MCS model based on the OLLA mechanism to simulate the PRB distortion induced by different channel conditions, thereby generalizing the data collected in our laboratory environment to the unknown communication environments that the target may inhabit.

\textbf{OLLA Random Walk Model.} For a given channel environment characterized by equilibrium $\bar{m}$, the OLLA mechanism confines the MCS to a tight cluster around $\bar{m}$ via a mean-reverting random walk. We formalize this mechanism as follows.

OLLA maintains an SINR offset $\delta_t$, updated per transmission by ACK/NACK outcome:
\begin{equation}
    \delta_{t+1} = \begin{cases}
        \delta_t + \Delta_{\text{ack}}, & \text{if ACK} \\
        \delta_t - \Delta_{\text{nack}}, & \text{if NACK}
    \end{cases}
\end{equation}
where $\Delta_{\text{nack}} / \Delta_{\text{ack}} = (1-p_{\text{target}})/p_{\text{target}}$ ensures long-run BLER converges to $p_{\text{target}}$. BLER follows a logistic (sigmoid) function of the gap $m_t - \bar{m}$. Therefore, selecting an MCS above $\bar{m}$ sharply increases NACK probability, pulling $\delta_t$ down; an MCS below $\bar{m}$ sees infrequent NACKs, allowing upward drift. Per-slot MCS is obtained by mapping adjusted SINR through 3GPP TS~38.214 table and clamping to $[\bar{m}-L, \bar{m}+L]$. Let $g(z)$ denote SINR-to-MCS mapping defined in TS~38.214, then
\begin{equation}
    \left\{\begin{aligned}
        m'_t &= g(\text{SINR}_{\bar{m}} + \delta_t) \\
        m_t   &= \max\big(\bar{m} - L,\; \min(\bar{m} + L,\, m'_t)\big)
    \end{aligned}\right.
\end{equation}

\begin{figure}[htbp]
    \centering
    \subfloat[Percentile profile overlay.]{\label{fig:eval-moment-a}\includegraphics[width=0.49\linewidth]{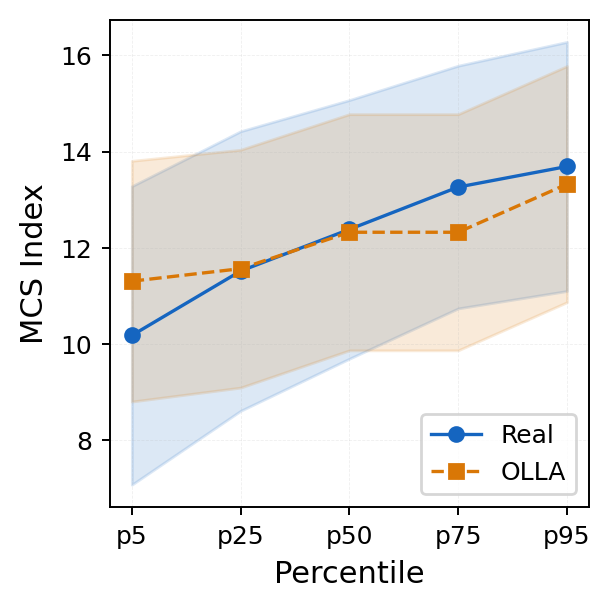}}
    \hfil
    \subfloat[MCS histograms.]{\label{fig:eval-moment-b}\includegraphics[width=0.49\linewidth]{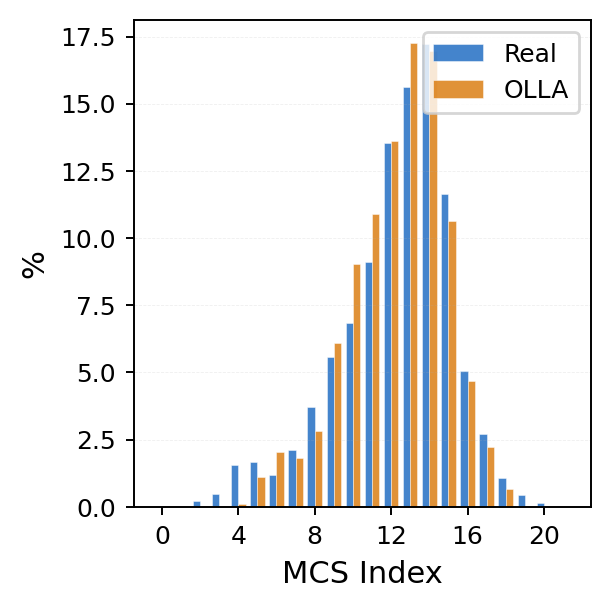}}

    \caption{Quantile preservation and histograms of the OLLA random walk model. (a)~Mean percentile profiles of real and synthetic MCS sequences, averaged over all windows with $\pm 1\sigma$ bands. (b)~Pooled MCS histograms over all windows.}\label{fig:eval-moment}
\end{figure}

At slow scale, we model $\bar{m}$ as a bounded random walk with step probability $1/\tau_{\text{drift}}$ estimated from empirical MCS autocorrelation decay. Figure~\ref{fig:eval-moment-a} validates that the OLLA random walk model faithfully reproduces real MCS dynamics. Panel (a) overlays percentile profiles of real and synthetic MCS sequences. The two profiles are nearly coincident across all five quantiles, with an RMSE of $0.68$ MCS indices. Figure~\ref{fig:eval-moment-b} overlays pooled MCS histograms, showing that the OLLA model faithfully reproduces the MCS distribution from the real trace.

Given this model, data augmentation proceeds as follows. For each IP trace collected in laboratory, we randomly sample a target equilibrium $\bar{m}_0$ uniformly from $[\hat{m}_{\text{base}}-L,\; \hat{m}_{\text{base}}+L]$, where $\hat{m}_{\text{base}}$ denotes mean MCS of laboratory collection environment, and run random walk to generate a synthetic MCS sequence $\{m_t\}$. Then, considering that MCS is not directly available in a real attack, we perform data augmentation by scaling RBs, which are directly observable from the spectrogram. We define \textit{maximum spectral contraction ratio} $\rho(m_t, m_{\text{lab}})$ as the ratio of PRBs required to carry a given TBS under sampled MCS $m_t$ versus laboratory MCS $m_{\text{lab}}$, computed directly from the TBS-to-PRB mapping specified in TS~38.214. For each TB in the laboratory trace, we scale its observed PRB count by $\rho$ to obtain synthetic PRB occupancy under target channel conditions. Repeating this procedure spans the operational MCS envelope without any additional physical RF collection.

\subsection{Cross-Modal Zero-Shot Fingerprinting}\label{sec:method-learning}
In prior work, pre-building fingerprints requires collecting and processing extensive spectrum data. However, once an attacker wishes to rebuild the fingerprint database, for instance to add a new application or update an existing fingerprint, they must re-establish the RF collection setup and re-collect over-the-air data. Consequently, existing air fingerprinting approaches are difficult to scale to larger datasets and lack operability, flexibility, and extensibility. We address this problem through cross-modal learning~\cite{DBLP:conf/cvpr/Duan0PLS025, DBLP:conf/iclr/DufumierNTT25}, which has been widely applied in natural language processing and vision to support LLM-based image generation. In our scenario, the attacker only needs to install a traffic capture application on a test mobile phone or Android emulator during the collection phase to synchronously obtain IP samples. However, such cross-modal fingerprint construction is non-trivial for two reasons:
\begin{itemize}
\item \textbf{Sparsity.} The spectrogram conveys far less information than IP sequences, at both the flow level and the packet level, so it is essential to extract the richest possible spectral information. 
\item \textbf{Variability.} The spectrogram exhibits greater variability across different channel environments, so the extracted features must be robust across channels.
\end{itemize}

\textbf{Feature representation.} Considering these potential challenges, we design a robust and expressive feature representation. As introduced in Section \ref{sec:channel_intro}, TBs observable in the spectrogram are aggregations of one or more IP packets, which entails information loss. Furthermore, PUCCH may carry NACK information, and although NACKs theoretically account for a small fraction, their prevalence can increase under poor channel conditions. Therefore, we choose IP data as the alignment anchor, since it can be pre-collected on the test UE side when the attacker builds the fingerprint database and is unaffected by varying physical-layer MCS conditions.

\begin{itemize}
\item \textbf{Spectral modality.} We extract a rich set of features from the spectrogram, including per-slot PUSCH carrier width, PUCCH activity count, and slot patterns of both channels. These features capture traffic morphology while being robust to MCS-induced distortion.
\item \textbf{IP modality.} We represent IP traffic collected via PcapDroid as packet-level features. To extract as many features as possible for more precise alignment, we collect both packet-level and flow-level features. Packet-level features include packet size, inter-arrival time, and protocol flags; flow-level features include flow duration and flow index. This multi-granularity representation captures both packet-level and flow-level patterns for cross-modal alignment.
\end{itemize}

\textbf{Backbone.} For each IP-spectral pair $\{(\mathcal{T}, \widetilde{W}_k)\}$, this yields paired training data $\{(\mathcal{T}, \widetilde{W}_k)\}$ where both modalities represent the same application activity. Two encoders encode $\{(\mathcal{T}, \widetilde{W}_k)\}$ to produce aligned embedding vectors. We adopt DF~\cite{DBLP:conf/ccs/SirinamIJW18} as the backbone for feature extraction on traffic representations. DF has demonstrated strong embedding capabilities and has been widely adopted in encrypted traffic classification~\cite{287133} and website fingerprinting~\cite{DBLP:journals/popets/RahmanSMG020, DBLP:conf/ccs/SirinamMR019}. We discard the final classification layer of DF and retain the 512-dimensional output of the penultimate layer as the traffic representation. We instantiate two independent DF backbones as encoders for two modalities. Let $\mathbf{x}^{\text{ip}} \in \mathbb{R}^{L_{\text{ip}}}$ denote normalized feature sequence of an IP trace $\mathcal{T}$, and $\mathbf{x}^{\text{spec}} \in \mathbb{R}^{L_{\text{spec}}}$ denote normalized spectral sequence of $\widetilde{W}$. Encoders produce $\ell_2$-normalized embeddings on a unit hypersphere:
\begin{equation}
\phi_{\text{ip}}(\mathcal{T}) = \frac{f_{\text{ip}}(\mathbf{x}^{\text{ip}})}{\|f_{\text{ip}}(\mathbf{x}^{\text{ip}})\|_2}, \quad
\phi_{\text{spec}}(\widetilde{W}) = \frac{f_{\text{spec}}(\mathbf{x}^{\text{spec}})}{\|f_{\text{spec}}(\mathbf{x}^{\text{spec}})\|_2}
\end{equation}
where $f_{\text{ip}}, f_{\text{spec}} \colon \mathbb{R}^{L} \to \mathbb{R}^{512}$ denote the DF backbone with independent parameters for IP and spectral modalities, respectively. 

\textbf{Contrastive Training Strategy.} Let $\phi_{\text{ip}}$ be an IP encoder that maps a traffic trace $\mathcal{T}$ to an embedding $\mathbf{h}^{\text{ip}}$, and $\phi_{\text{spec}}$ be a spectral encoder that maps a spectral traffic profile $\widetilde{W}[t]$ to an embedding $\mathbf{h}^{\text{spec}}$. We need to align intra-modal distances within each modality, increase inter-class separation, and pull embeddings of different modalities closer together. We address these requirements through a composite loss function.

\textbf{InfoNCE loss.} We treat each IP-spectral pair as a positive sample and all other embeddings in the batch, regardless of modality, as negatives. Let $\mathbf{h}_i$ denote the embedding of sample $i$, drawn from the union of IP and spectral embeddings within a minibatch. InfoNCE loss for a positive pair $(i, j)$ is defined as
\begin{equation}
    \mathcal{L}_{\text{InfoNCE}} = -\log \frac{\exp(\mathbf{h}^{\text{ip}}_i \cdot \mathbf{h}^{\text{spec}}_i / \tau)}{\sum_{j} \exp(\mathbf{h}^{\text{ip}}_i \cdot \mathbf{h}^{\text{spec}}_j / \tau)}
\end{equation}
where $\tau$ is a temperature hyperparameter and the denominator sums over all spectral embeddings in the batch. This loss encourages embeddings of the same application to be close in shared space while pushing different applications apart.

\textbf{SupCon loss.} To further enhance inter-class separation, we add a supervised contrastive loss that explicitly clusters same-class embeddings and separates different-class embeddings across both modalities. Let $\mathbf{h}_i$ denote any embedding from $\{\mathbf{h}^{\text{ip}}\} \cup \{\mathbf{h}^{\text{spec}}\}$, and let $P(i)$ be set of indices of samples sharing same application label as sample $i$ within batch. SupCon loss is defined as
\begin{equation}
    \mathcal{L}_{\text{SupCon}} = -\sum_{i} \frac{1}{|P(i)|} \sum_{p \in P(i)} \log \frac{\exp(\mathbf{h}_i \cdot \mathbf{h}_p / \tau)}{\sum_{a \neq i} \exp(\mathbf{h}_i \cdot \mathbf{h}_a / \tau)}
\end{equation}
where the denominator sums over all other samples in the batch. This loss encourages tighter intra-class clustering and wider inter-class separation, improving discriminability of embedding space.

\textbf{Align loss.} We directly minimize the cosine distance between paired IP and spectral embeddings to pull the two modalities closer. Since both encoders produce $\ell_2$-normalized embeddings, the cosine loss simplifies to
\begin{equation}
    \mathcal{L}_{\text{align}} = 1 - \mathbf{h}^{\text{ip}} \cdot \mathbf{h}^{\text{spec}}
\end{equation}
minimized over all augmented pairs $\{(\mathcal{T}_i, \widetilde{W}_{i,k})\}$. As the same IP trace generates multiple spectral profiles under different MCS values, this loss drives the spectral encoder to collapse channel-induced variation into a consistent embedding anchored by the IP trace.

The full training objective combines cross-modal alignment, intra-modal inter-class separation, and pairwise cross-modal attraction:
\begin{equation}
    \mathcal{L} = \mathcal{L}_{\text{InfoNCE}} + \alpha \mathcal{L}_{\text{SupCon}} + \beta \mathcal{L}_{\text{align}}
\end{equation}
where $\alpha$ and $\beta$ balance the three terms. Minimizing $\mathcal{L}_{\text{InfoNCE}}$ aligns the two modalities in a shared embedding space and provides coarse inter-class separation, $\mathcal{L}_{\text{SupCon}}$ further sharpens class boundaries across both modalities, and $\mathcal{L}_{\text{align}}$ enforces fine-grained pairwise proximity between each IP-spectral pair.

\textbf{Zero-shot classification via IP anchoring.} Aligned space enables a practical zero-shot attack. For a target application whose IP traffic the attacker can capture in the training phase, IP encoder $\phi_{\text{ip}}$ produces a reference embedding $\mathbf{h}^{\text{ip}}_{\text{ref}}$ without requiring any physical RF measurement. A physically observed spectral traffic profile $W[t]$ is encoded via $\phi_{\text{spec}}$ and classified by nearest-neighbor matching against a gallery of IP-derived reference embeddings. Since all embeddings are $\ell_2$-normalized, we use cosine distance as the matching metric:
\begin{equation}
    \hat{y} = \arg\min_{c \in \mathcal{C}} \; \bigl(1 - \phi_{\text{spec}}(W[t]) \cdot \mathbf{h}^{\text{ip}}_{\text{ref}, c}\bigr)
\end{equation}
where $\mathcal{C}$ is set of candidate applications. For applications whose IP traces were used during alignment training, classification requires zero physical samples. When an attacker needs to enroll a newly interested application or update an existing application version, they only need to collect a single IP trace from a test UE or Android emulator, compute its reference embedding, and add it to the reference gallery. A single IP trace, captured at negligible cost, replaces dedicated RF collection campaigns across multiple MCS conditions.

\section{Evaluation}
We present the evaluation of \name{}  through comprehensive real-world experiments, demonstrating its feasibility.

\begin{itemize}
\item \textbf{Implementation.} We prototype \name{} using Torch 2.1.0 and Python 3.8. We execute our attack on a server running Ubuntu 22.04 LTS, equipped with a single NVIDIA A40 24GB GPU. Following prior works, we split the dataset into training, validation, and testing, with a ratio of 8:1:1, ensuring no overlap across splits. For cross-domain zero-shot attacks, we use IP traffic as the anchor and evaluate using completely disjoint and unseen spectra at attack time. DF is implemented using official code with default settings.


\item \textbf{Testbed.} We follow prior work~\cite{11034666, 10.5555/3766078.3766355, yoon2024scalablerobustmobileactivity} in using open5GS~\cite{open5gs} and srsRAN~\cite{srsRAN} to establish a 5G NR BS for controlled data collection. We use a USRP B210 as the sniffer, capturing signals at a 20\,MHz bandwidth. For public privacy and ethical considerations, all the experiments are strictly controlled.

\item \textbf{Baseline:} We choose PTTF \cite{11034666} as the baseline. PTTF conducts cellular network fingerprint attacks by collecting uplink spectrum data, which is consistent with our scenario. It designs a random forest classifier for classifying apps and services.

\item \textbf{Dataset:} We select the most popular (based on download times) apps as targets for identification. All apps are the latest versions in the App Store and span diverse categories of daily usage. The collection was completed within two weeks and covers 78 apps in total, representing the largest-scale implementation in the air fingerprinting domain to date. After filtering out failed apps due to network connectivity issues and root environment detection, we obtained over 9k valid samples. In addition, we recollected over 9k cross-environment test samples under four different channel conditions.
\end{itemize}

\subsection{Side-channel Availability}\label{sec:eval-sidechannel}
We first conduct a qualitative analysis of the key observations presented in Section~\ref{sec:motiv} to evaluate whether the proposed side channel remains effective across diverse channel conditions.

We use a Xiaomi~12 smartphone as the UE, running a live streaming application to generate realistic bidirectional traffic. The UE connects to a 5G NR BS~\cite{srsRAN}. To emulate diverse channel conditions, we adjust the UE-to-BS distance and introduce obstructions such that the \texttt{ul\_mcs} reported in the srsRAN console log fluctuates around target values while keeping all other BS settings at their defaults. We collect data under three representative channel conditions spanning the operational range from poor to excellent channel quality (mean UL MCS $\mu_{\text{UL}} \in \{1.4, 14.2, 26.8\}$). The downlink MCS is determined adaptively by the BS scheduler and does not follow the uplink MCS. For each condition, we simultaneously capture two data streams: (i)~a pcap trace of the UE's IP traffic, and (ii)~the srsRAN MAC-layer log containing per-slot PUSCH and PUCCH transmission records. We then quantify their similarity using two complementary metrics:\textbf{Histogram Overlap} (HistOv = $\sum_i \min(H_x[i], H_y[i])$) and \textbf{Kolmogorov--Smirnov (KS) Statistic = $\sup_x |F_x(x) - F_y(x)|$}.

\begin{figure}[htbp]
\centering
\includegraphics[width=\columnwidth]{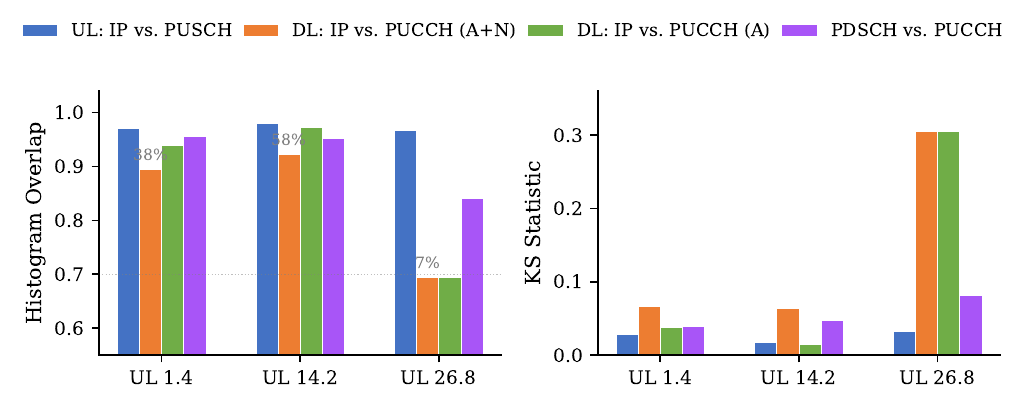}
\caption{Side channel availability across three channel conditions. Three comparisons are shown per condition: uplink IP vs.\ PUSCH, downlink IP vs.\ PUCCH (ACK+NACK), downlink IP vs.\ PUCCH (ACK only) and PUCCH vs.\ PDSCH.}\label{fig:sidechannel-fidelity}
\end{figure}

Figure~\ref{fig:sidechannel-fidelity} reports the results. We conduct four pairwise comparisons: (1)~IP uplink packets vs.\ PUSCH transmissions, validating the core uplink side channel; (2)~IP downlink packets vs.\ PUCCH Format~1 activity (ACK+NACK), evaluating the raw downlink proxy; (3)~IP downlink packets vs.\ PUCCH ACK-only (NACK removed), quantifying the noise introduced by HARQ retransmission feedback; and (4)~DL PDSCH transmissions vs.\ PUCCH Format~1 activity, validating that PUCCH feedback faithfully reflects the actual downlink data channel behavior.

\textit{IP uplink vs.\ PUSCH.} The IP uplink vs.\ PUSCH comparison exhibits consistently strong similarity across all three conditions: HistOv ranges from $0.968$ to $0.989$ (mean $0.977 \pm 0.008$), and KS statistics remain below $0.034$ (mean $0.022 \pm 0.010$). This means that a passive adversary observing only the per-slot PUSCH activity sequence can reconstruct the uplink IP traffic pattern with super-high similarity without decoding any control-plane message. Critically, this similarity is invariant to the channel conditions. The maximum HistOv deviation across conditions spanning UL MCS from $1.4$ to $26.8$ is only $0.021$, confirming that the side channel remains structurally stable across the operational MCS envelope.

\textit{IP downlink vs.\ PUCCH.} The IP downlink vs.\ PUCCH comparison reveals two effects. First, PUCCH activity provides a viable downlink proxy despite the downlink MCS varying independently. HistOv exceeds $0.89$ for two of three conditions, demonstrating that HARQ feedback density indeed reflects downlink traffic intensity even without access to the downlink spectrum. Second, NACK transmissions act as a systematic confounder. Removing NACKs and retaining only ACK bits improves HistOv by $0.044$--$0.049$ across the two conditions with non-negligible NACK rates, and reduces the KS statistic by $0.029$--$0.049$. At UL MCS~$=26.8$ where only $7.2\%$ of HARQ bits are NACKs, removing them produces no measurable change. We emphasize that ACK and NACK transmissions are spectrally indistinguishable since both occupy a single RB at the band edge and produce identical energy signatures in the spectrogram. Consequently, methods such as PTTF~\cite{11034666} cannot separate ACK from NACK in practice, which inherently limits their precision.

\textit{Abnormal phenomenon.} We observe a counterintuitive pattern: At UL MCS~$=26.8$ (DL MCS $=20.4$), the DL similarity drops sharply (HistOv $=0.695$, KS $=0.305$) regardless of NACK filtering. Two compounding effects explain this result. First, the live streaming application adapts its video bitrate to the available bandwidth, generating substantially more downlink IP packets under favorable channel conditions.  Each downlink TB carries a large payload, aggregating multiple IP packets into a single TB. Second, at high data rates the UE multiplexes HARQ-ACK bits for multiple PDSCH transmissions into a single HARQ-ACK codebook~\cite{3GPP_TS_38_213, 3GPP_TS_38_212} rather than transmitting a separate PUCCH per downlink TB. Nevertheless, as Figure~\ref{fig:sidechannel-fidelity} shows, PUSCH similarity remains high across all conditions. This sustained uplink fidelity underpins the precision advantage of \name{} over PTTF.

\subsection{Performance Comparison}
Since PTTF does not support few-shot classification, we compare the two methods on their ability to classify all 78 applications using uplink physical-layer data under a standard supervised setting. The dataset is split 8:1:1 for training, validation, and testing. We report accuracy, macro-averaged precision, recall, and F1-score for both in-domain and cross-channel evaluation. For DF with augmentation, we report the result per channel condition achieved across the four augmented models. Specifically, we train on data collected under the best channel condition and evaluate on data collected under different channel conditions unseen during training.

\begin{figure}[htbp]
\centering
\includegraphics[width=\columnwidth]{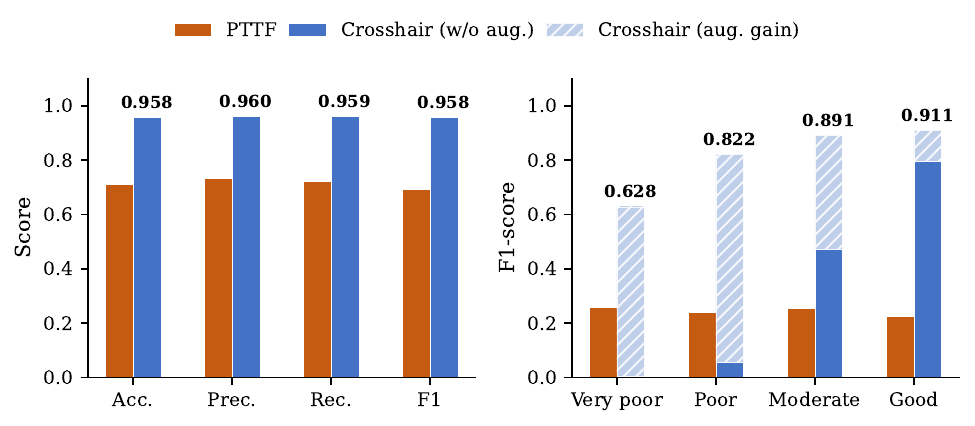}
\caption{Classification performance comparison between PTTF and \name{}. (a)~In-domain results across four metrics. (b)~Cross-channel F1-score, where each DF bar is a stacked bar.}
\label{fig:cls-comparison}
\end{figure}

Figure~\ref{fig:cls-comparison}(a) reports the in-domain comparison across four metrics. DF consistently achieves scores above 0.95 across all four metrics, while PTTF ranges from 0.69 to 0.73. The gap stems from two factors. First, \name{} learns more discriminative and channel-robust representations by reconstructing bidirectional traffic from the uplink spectrum, whereas PTTF uses handcrafted statistical features fed into a random forest classifier. Second, PTTF was originally evaluated on only 20 applications. Our evaluation covers 78 applications, nearly four times that scale, which exceeds the classification capacity of a random forest model operating on shallow features.

Figure~\ref{fig:cls-comparison}(b) reports the cross-channel F1-score. PTTF remains flat around 0.22 to 0.26 regardless of channel condition, indicating that its handcrafted features are insensitive to channel-induced distortion but lack discriminative power. This is consistent with our finding in Figure~\ref{fig:sidechannel-fidelity} that the PUCCH-to-downlink-IP similarity measured under the best channel condition differs substantially from that observed under other channel conditions. The DF stacked bars reveal two effects. The lower solid portion shows that without augmentation, DF degrades severely under the very poor and poor conditions (F1 below 0.06), confirming that the convolutional features are tightly coupled to the training channel regime. The upper hatched portion quantifies the recovery from augmentation, adding 0.62 and 0.77 under the very poor and poor conditions, and 0.42 and 0.11 under the moderate and good conditions. With augmentation, DF reaches 0.63 under the very poor condition and surpasses 0.89 under the moderate and good conditions, exceeding PTTF's best result by a wide margin. These substantial gains from data augmentation validate the effectiveness of the proposed OLLA random walk model and spectral warping method.

\subsection{Zero-shot Classification}

We evaluate the cross-modal zero-shot classification performance of \name{} under in-domain condition, where training and testing data are drawn from the same channel environment. We train on the best channel environment with 78 applications, partitioned into 62 training applications and 16 validation applications. The validation applications are entirely disjoint from the training set.

We adopt a few-shot evaluation. In each episode, $k$ support samples per class are provided ($k \in \{1, 5, 10, 20\}$). The query samples are classified by cosine nearest-neighbor matching against the support prototypes. We report three metrics. IP-only uses the IP encoder $\phi_{\text{ip}}$ to encode both support and query samples within the IP modality. Air-only uses the spectral encoder $\phi_{\text{spec}}$ to encode support and query samples within the spectral modality, simulating the attack scenario in which the attacker has previously collected a few spectral samples to enroll applications in a \textit{few-shot} setting. Cross-modal zero-shot uses IP-derived prototypes as support and spectral embeddings as queries, simulating the attack scenario in which the attacker classifies previously unseen physical spectrum in a cross-domain \textit{zero-shot} setting.

\begin{table}[htbp]
\centering
\caption{Accuracy of Few-shot and zero-shot classification. Training and testing are both conducted under the best channel environment.}
\label{tab:fewshot-main}
\small
\begin{tabular}{lccc}
\hline
\textbf{Shot} & \textbf{IP-only} & \textbf{Air-only (few-shot)} & \textbf{Cross-modal} \\
\hline
1  & 0.9009 & 0.8831 & 0.8286 \\
5  & 0.9618 & 0.9538 & 0.9019 \\
10 & 0.9665 & 0.9598 & 0.9139 \\
20 & 0.9692 & 0.9609 & \textbf{0.9223} \\
\hline
\end{tabular}
\end{table}

Table~\ref{tab:fewshot-main} reports the results. In the cross-modal zero-shot setting, \name{} achieves 0.8286 accuracy at 1-shot and 0.9143 at 20-shot. Notably, with only 5 IP samples the attacker achieves over 90\% spectrum classification accuracy, despite having never observed the target spectrum before. This confirms that the embedding space learned by \name{} effectively aligns the two modalities. In the Air-only few-shot setting, which simulates the scenario where the attacker has previously collected a few spectral samples to enroll applications, \name{} achieves 0.8831 at 1-shot and 0.9609 at 20-shot. The Air-only accuracy is consistently higher than the cross-modal accuracy. This gap reflects the additional difficulty of aligning IP and spectral modalities due to inherent information loss across modalities. Nevertheless, the cross-modal accuracy at 1-shot (0.8286) is only 5.5 points below the Air-only accuracy at 1-shot (0.8831), and the cross-modal accuracy at 20-shot (0.9223) approaches the Air-only ceiling. The IP-only baseline reaches 0.9692 at 20-shot, serving as the upper bound since IP data is unaffected by physical-layer distortions. These results demonstrate that \name{} learns a semantically meaningful embedding space in which applications can be recognized across modalities without laborious physical RF enrollment, at only a modest cost relative to single-modality alternatives.

\subsection{Cross-Channel Evaluation}\label{sec:eval-cross}

The in-domain results above assume training and testing under the same channel condition. In practice, the attacker cannot control or predict the target user's channel environment. To evaluate \name{} under realistic cross-channel conditions, we adjust the channel environment by varying the distance between the UE and the BS, introducing physical obstructions, which collectively drive the MCS to fluctuate across distinct regimes. We consider four representative conditions spanning the operational range, referred to as \textit{very poor}, \textit{poor}, \textit{moderate}, and \textit{good} in decreasing order of signal degradation. The training data is collected under the best channel condition. 

We evaluate \name{} with the cross-channel data augmentation module disabled to establish a baseline. We then enable augmentation from the training dataset collected from the best channel condition to unknown channel conditions using the method described in Section~\ref{sec:method-twin}. For each evaluation condition, we report cross-modal accuracy achieved across the four augmented models.

\begin{table}[htbp]
\centering
\caption{Cross-channel classification accuracy. Training is conducted with the dataset collected from the best channel conditions.}
\label{tab:cross-channel}
\small
\begin{tabular}{lcccc}
\hline
\textbf{Channel} & \textbf{IP-only} & \textbf{Cross-modal} & \textbf{Cross-modal} \\
& & \textbf{(w/o aug.)} & \textbf{(with aug.)}\\
\hline
Very poor  & 0.9533 & 0.2614 & \textbf{0.6014}(+0.3400) \\
Poor       & 0.9386 & 0.3280 & \textbf{0.8600}(+0.5320) \\
Moderate   & 0.9541 & 0.5713 & \textbf{0.9435}(+0.3722) \\
Good       & 0.9468 & 0.7684 & \textbf{0.9656}(+0.1972) \\
\hline
\end{tabular}
\end{table}

Table~\ref{tab:cross-channel} reports the results. Without augmentation, the cross-modal accuracy degrades sharply as the channel condition worsens, dropping from 0.7684 under the good condition to 0.2614 under the very poor condition, a decline of over 50 percentage points. This confirms that PRB occupancy patterns undergo substantial distortion across channel regimes, and a model trained exclusively under one condition fails to generalize. With cross-channel data augmentation enabled, the cross-modal accuracy improves substantially across all four conditions. Notably, under most commonly encountered channel conditions, cross-modal zero-shot recognition still exceeds 85\% accuracy. The gains under the very poor condition are less pronounced, because we observe that a large fraction of RB allocations exceed the bandwidth allocation limit. In practice, this would cause stalling or trigger a handover to a stronger BS~\cite{DBLP:journals/ejwcn/HaghrahAAN23, 7510709}. These results demonstrate the effectiveness of cross-channel data augmentation in bridging the generalization gap across channel environments.

\subsection{Ablation Study}

We conduct an ablation study to quantify the contribution of each loss component in the contrastive training strategy described in Section~\ref{sec:method-learning}. Starting from the SupCon baseline, which provides supervised inter-class separation within each modality, we incrementally add InfoNCE for cross-modal contrastive alignment and the cosine similarity loss for pairwise cross-modal attraction. All experiments use the DF backbone without data augmentation.

\begin{table}[htbp]
\centering
\caption{Ablation study of loss components.}
\label{tab:ablation}
\small
\begin{tabular}{lcc}
\hline
\textbf{Configuration} & \textbf{Cos-sim} & \textbf{Cross-modal Acc.} \\
\hline
SupCon only              & 0.2636  & 0.8015 \\
SupCon + InfoNCE         & 0.3476  & 0.8602 \\
SupCon + Align           & 0.6473  & 0.8923 \\
SupCon + InfoNCE + Align & 0.7350 & \textbf{0.9223} \\
\hline
\end{tabular}
\end{table}

Table~\ref{tab:ablation} reports the results. The SupCon-only baseline achieves 0.8015 cross-modal accuracy, demonstrating that supervised contrastive learning alone provides a reasonable embedding space but insufficient cross-modal alignment. Adding InfoNCE raises the accuracy from 0.8015 to 0.8602 by introducing cross-modal negative pairs that explicitly align the IP and spectral modalities. Adding Align loss to SupCon yields a larger gain, raising the accuracy to 0.8923. The Align loss directly minimizes the pairwise distance between each IP-spectral pair, which is more effective at pulling the two modalities together than the batch-level contrastive signal of InfoNCE alone. The full combination of all three losses achieves the highest performance across all metrics, with a best accuracy of 0.9223. Removing any component degrades performance, confirming that all three losses contribute non-redundant signals to the training objective.

\section{Discussion}
\subsection{Limitations}
While our spectrogram-based app fingerprinting attack demonstrates high success rates, several limitations constrain its applicability and generalizability.

\textbf{Distinguishing UEs:} 
In typical deployment scenarios, an attacker can effectively suppress interference originating from other UEs. By strategically adjusting antenna orientation and receiver placement, the attacker can ensure that the target UE's signal remains either exclusive or dominant within the observed spectrum. The use of advanced receiving hardware (e.g., beamforming antennas) further enhances signal acquisition quality. In multi-user scenarios, attackers can deploy multiple sniffers to obtain signals from the spectrum in a manner similar to triangulation~\cite{10.1145/3636534.3690709, 11034666}. Actually, distinguishing individual UEs pertains to the domain of RF fingerprinting. This area has been extensively studied in prior work~\cite{DBLP:journals/tifs/YangL24, DBLP:journals/corr/abs-2512-10809, DBLP:journals/iotj/PengPFL24,DBLP:journals/wcl/JangKKMKK25, 11003929}, and we regard RF fingerprinting as an orthogonal area to our work. We envision conducting future research in this field to complement the scope of \name{}, rather than competing with it.

\textbf{MCS stability.} Our side channel is based on the empirical observation that MCS remains stable when users freely operate their phones, even with natural changes in device position. We designed data augmentation to match varying MCS conditions across different channels. However, we did not evaluate rapidly moving users, as this scenario falls outside the realistic threat model. An attacker deploying a USRP at a fixed location cannot feasibly track a UE in real time. Intuitively, it is difficult for an attacker to track a moving vehicle with an RF sniffer, since the phone quickly exits the sniffing range.

\textbf{Generalizability:} To facilitate parameter tuning and protect public privacy, our evaluation was conducted in a controlled laboratory environment, which may not fully capture the complexity and variability of real-world scenarios. Environmental noise and user mobility can degrade the quality of the over-the-air signals that the attacker can capture, thereby affecting attack performance. We do not consider scenarios where air signals are unavailable, as already stated in the threat model, since constructing air fingerprints to launch an attack is inherently difficult under such conditions. Owing to the cross-modal zero-shot design, we believe our method possesses strong scalability.

\subsection{Countermeasures}
Several approaches may defend against this attack, yet each introduces distinct trade-offs.

\textbf{MCS randomization.} By randomizing MCS selection, the attacker can no longer reliably map PRB occupancy back to IP packet length, thereby disrupting the uplink side channel. However, MCS randomization makes spectral efficiency extremely poor, severely degrading the experience of all cellular users since spectrum is a finite resource. Moreover, the 3GPP OLLA mechanism (TS 38.214) is explicitly designed to maximize link adaptation efficiency, and randomizing MCS is tantamount to voluntarily forfeiting the gains of link adaptation, contradicting the original intent of the protocol design.

\textbf{PRB regularization.} When allocating uplink PUSCH resources, the BS scheduler deliberately pads the PRB allocation to a fixed set of discrete values, such that PUSCH always appears as one of several predefined bandwidths in the spectrogram regardless of the actual TBS. This is analogous to packet-length padding strategies in traditional network traffic analysis defenses. Specifically, the BS can quantize PRB allocation into K levels (e.g., $\{2, 4, 8, 16, 32, \ldots\}$ PRBs) and round up to the nearest available level at each scheduling instance. Nevertheless, this defense still wastes valuable spectrum resources. In addition, it requires modifications to the existing 5G scheduler.

\section{Related Works}
A number of studies have demonstrated the feasibility of eavesdropping on users’ private information by exploiting leaked information over the air interface. 

\textbf{Cellular Network Air Attack.} Kotuliak et al. \cite{DBLP:conf/uss/KotuliakELRC22} introduce a new type of IMSI Catcher that can extract a device's IMSI and bind it to its current Temporary Mobile Subscriber Identity (TMSI). Rupprecht et al. \cite{8835335} present an identity-mapping attack that matches RNTI to TMSI, enabling us to identify users within a cell and serving as a stepping stone for follow-up attacks. Furthermore, they demonstrate the feasibility of constructing a fingerprint by collecting air traffic data. Hong et al. \cite{hong2026passive} correlate time-stamped visual observations of device usage with captured air messages, thereby determining the Globally Unique Temporary Identifier (GUTI) of each device within the camera's field of view. However, the aforementioned efforts in identity tracking have become irrelevant with the more secure design of 5G NR. 5G SUCI-Catcher \cite{10.1145/3448300.3467826} captures the encrypted SUCI and queries a certain SUCI to determine the existence of the user. In particular, the 3GPP considers the attacks less powerful compared to IMSI-Catching. Existing 5G sniffers \cite{10.1145/3680121.3697808, 10.5555/3766078.3766355} require users to initiate the access process either consciously or unconsciously. Therefore, attackers must sniff at restricted locations (e.g., airports) or deploy additional risky positive attacks. PTTF \cite{11034666} proposes a 5G NR fingerprint attack over the air interface to identify specific app activity by using ACK/NACK information. However, the ACK/NACK information on the PUCCH may be absent during uplink data transmission.


\textbf{Air Side Channel Attack.} LeakyBeam \cite{DBLP:conf/ndss/0002CH0H25} presents a side channel that utilizes beamforming feedback information (BFI) to infer the user's movement. AppListener \cite{DBLP:conf/uss/NiL0ZX23} leverages RF energy harvesting as a side-channel to eavesdrop on the activities of a WiFi-connected smartphone app. Unlike these works on WiFi Downlink sniffing, \name{} exploits a new side channel in cellular network Uplink data transmission, which is practical and challenging. Lakshmanan et al. \cite{DBLP:conf/uss/LakshmananBKCH21} utilize Carrier Aggregation (CA) information leaked through the air interface to achieve passive location tracking of targeted cellular user devices. Jederman et al. \cite{DBLP:conf/uss/JedermannSLS24} exploit interaction information observed passively from the downlink of wandering communication satellites over wide beams to estimate the region in which users are located. Jawne et al. \cite{jawne2025aiassisted} demonstrate the critical role of spectrograms in capturing distinct hardware impairments to identify rogue 5G devices. Unlike these works, \name{} leverages the harvested RF spectrograms to pinpoint fine-grained in-app activities.

\section{Conclusion}

This paper presents \name{}, a passive application fingerprinting attack on 5G NR that exploits a physical-layer side channel in the uplink spectrum. We show that under the stable MCS induced by 3GPP-mandated link adaptation, the PRB occupancy observable in the spectrogram reliably reflects the IP packet length, enabling bidirectional traffic reconstruction from uplink spectrum capture alone without decoding any message. Based on this side channel, \name{} employs a three-stage pipeline for an air fingerprinting attack. Extensive experiments on a 5G NR testbed with commercial UEs demonstrate that \name{} achieves over 90\% application recognition accuracy. Furthermore, \name{} outperforms the existing physical-layer fingerprinting method across diverse MCS regimes and channel conditions. We hope this work spurs further research into uplink-aware countermeasures and inspires the standardization of physical-layer privacy protections in future 3GPP releases.

\bibliographystyle{IEEEtran}
\bibliography{bib/sample-base}

@String{Computing = "Computing" }

@String{Computer = "{IEEE} Computer" }

@String{Springer = "Springer-Verlag" }

@inproceedings{11145,
author = {Kohls, Katharina and Rupprecht, David and Holz, Thorsten and P\"{o}pper, Christina},
title = {Lost traffic encryption: fingerprinting LTE/4G traffic on layer two},
year = {2019},
doi = {10.1145/3317549.3323416},
booktitle = {ACM Conference on Security and Privacy in Wireless and Mobile Networks(WiSec)},
pages = {249–260}
}

@inproceedings{8835335,
  author       = {David Rupprecht and
                  Katharina Kohls and
                  Thorsten Holz and
                  Christina P{\"{o}}pper},
  title        = {Breaking {LTE} on Layer Two},
  booktitle    = {2019 {IEEE} Symposium on Security and Privacy, {SP} 2019, San Francisco,
                  CA, USA, May 19-23, 2019},
  pages        = {1121--1136},
  publisher    = {{IEEE}},
  year         = {2019},
  doi          = {10.1109/SP.2019.00006}
}

@ARTICLE{11034666,
  author={Zhang, Hang and Wei, Dong and Jiang, Nan and Zhang, Meng and Meng, Xiang and Yang, Yang and Huang, Weiqing},
  journal={IEEE Transactions on Information Forensics and Security}, 
  title={Passive Multi-User Traffic Analysis Based on 5G NR/LTE Physical Layer}, 
  year={2025},
  volume={20},
  number={},
  pages={6794-6809},
  doi={10.1109/TIFS.2025.3578917}
  }

@inproceedings{bae:2022:watching,
  author       = {Sangwook Bae and
                  Mincheol Son and
                  Dongkwan Kim and
                  CheolJun Park and
                  Jiho Lee and
                  Sooel Son and
                  Yongdae Kim},
  editor       = {Kevin R. B. Butler and
                  Kurt Thomas},
  title        = {Watching the Watchers: Practical Video Identification Attack in {LTE}
                  Networks},
  booktitle    = {31st {USENIX} Security Symposium, {USENIX} Security 2022, Boston,
                  MA, USA, August 10-12, 2022},
  pages        = {1307--1324},
  publisher    = {{USENIX} Association},
  year         = {2022}
}

@INPROCEEDINGS{10202613,
  author={Baek, Jaejong and Soundrapandian, Pradeep Kumar Duraisamy and Kyung, Sukwha and Wang, Ruoyu and Shoshitaishvili, Yan and Doupé, Adam and Ahn, Gail-Joon},
  booktitle={2023 53rd Annual IEEE/IFIP International Conference on Dependable Systems and Networks (DSN)}, 
  title={Targeted Privacy Attacks by Fingerprinting Mobile Apps in LTE Radio Layer}, 
  year={2023},
  volume={},
  number={},
  pages={261-273},
  doi={10.1109/DSN58367.2023.00035}}

@INPROCEEDINGS{10723461,
  author={Islam, Md Ruman and Anwar, Raja Hasnain and Mastorakis, Spyridon and Raza, Muhammad Taqi},
  booktitle={2024 IEEE 21st International Conference on Mobile Ad-Hoc and Smart Systems (MASS)}, 
  title={Characterizing Encrypted Application Traffic Through Cellular Radio Interface Protocol}, 
  year={2024},
  volume={},
  number={},
  pages={321-329},
  doi={10.1109/MASS62177.2024.00050}}

@InProceedings{10.1007/978-981-96-9872-1_4,
author="Zhang, Wenao
and Chen, Shuhui
and Wei, Ziling
and Zhang, Xinyu
and Xing, Qianqian
and Su, Jinshu",
editor="Huang, De-Shuang
and Chen, Wei
and Pan, Yijie
and Chen, Haiming",
title="Cellular-Snooper: A General and Real-Time Mobile Application Fingerprinting Attack in LTE Networks",
booktitle="Advanced Intelligent Computing Technology and Applications",
year="2025",
publisher="Springer Nature Singapore",
address="Singapore",
pages="39--53",
isbn="978-981-96-9872-1"
}

@article{Balasingam2017PosterBL,
  title={Poster: Broadcast LTE Data Reveals Application Type},
  author={Arjun Balasingam and Manu Bansal and Rakesh Misra and Rahul Tandra and Aaron Schulman and Sachin Katti},
  journal={Proceedings of the 23rd Annual International Conference on Mobile Computing and Networking},
  year={2017}
}

@INPROCEEDINGS{10181886,
  author={Chen, Yongming and Tong, Yuzhou and Hwee, Gwee Bah and Cao, Qi and Razul, Sirajudeen Gulam and Lin, Zhiping},
  booktitle={2023 IEEE International Symposium on Circuits and Systems (ISCAS)}, 
  title={Real-time Traffic Classification in Encrypted Wireless Communication Network}, 
  year={2023},
  volume={},
  number={},
  pages={1-5},
  doi={10.1109/ISCAS46773.2023.10181886}}

@INPROCEEDINGS{11304648,
  author={Zhang, Wenao and Chen, Shuhui and Liao, Junhong and Wei, Ziling and Gong, Mengyi},
  booktitle={2025 IEEE International Performance, Computing, and Communications Conference (IPCCC)}, 
  title={I Know Who You are: An Identity Mapping Attack Based on Time Series Similarity in Mobile Networks}, 
  year={2025},
  volume={},
  number={},
  pages={1-9},
  doi={10.1109/IPCCC66453.2025.11304648}}

@article{9210554,
  author       = {Hoang Duy Trinh and
                  {\'{A}}ngel Fern{\'{a}}ndez Gamb{\'{\i}}n and
                  Lorenza Giupponi and
                  Michele Rossi and
                  Paolo Dini},
  title        = {Mobile Traffic Classification Through Physical Control Channel Fingerprinting:
                  {A} Deep Learning Approach},
  journal      = {{IEEE} Trans. Netw. Serv. Manag.},
  volume       = {18},
  number       = {2},
  pages        = {1946--1961},
  year         = {2021},
  doi          = {10.1109/TNSM.2020.3028197}
}

@INPROCEEDINGS{9631415,
  author={Zhai, Liuqun and Qiao, Zhuang and Wang, Zhongfang and Wei, Dong},
  booktitle={2021 IEEE Symposium on Computers and Communications (ISCC)}, 
  title={Identify What You are Doing: Smartphone Apps Fingerprinting on Cellular Network Traffic}, 
  year={2021},
  volume={},
  number={},
  pages={1-7},
  doi={10.1109/ISCC53001.2021.9631415}}

@INPROCEEDINGS{11047734,
  author={Mei, Wenming and Meng, Xiang and He, Chen and Qiu, Lanxin and Zhou, Yihan and Tang, Yize and Ji, Rong and Zhang, Hang},
  booktitle={2025 5th International Conference on Artificial Intelligence and Industrial Technology Applications (AIITA)}, 
  title={Passive Traffic Analysis for Instant Messaging Applications in Mobile Communication Networks Air Interface}, 
  year={2025},
  volume={},
  number={},
  pages={693-698},
  doi={10.1109/AIITA65135.2025.11047734}}

@inproceedings{10.1145/3636534.3690709,
author = {Oh, Taekkyung and Bae, Sangwook and Ahn, Junho and Lee, Yonghwa and Hoang, Tuan Dinh and Kang, Min Suk and Tippenhauer, Nils Ole and Kim, Yongdae},
title = {Enabling Physical Localization of Uncooperative Cellular Devices},
year = {2024},
isbn = {9798400704895},
publisher = {Association for Computing Machinery},
address = {New York, NY, USA},
doi = {10.1145/3636534.3690709},
booktitle = {Proceedings of the 30th Annual International Conference on Mobile Computing and Networking},
pages = {1530–1544},
numpages = {15},
location = {Washington D.C., DC, USA},
series = {ACM MobiCom '24}
}

@ARTICLE{11003929,
  author={Zhang, Junqing and Ardizzon, Francesco and Piana, Mattia and Shen, Guanxiong and Tomasin, Stefano},
  journal={IEEE Transactions on Information Forensics and Security}, 
  title={Physical Layer-Based Device Fingerprinting for Wireless Security: From Theory to Practice}, 
  year={2025},
  volume={20},
  number={},
  pages={5296-5325},
  doi={10.1109/TIFS.2025.3570118}}

@INPROCEEDINGS{10757717,
  author={Aphayavong, Bounlhom and Fei, Xiao and Lu, Jialiang and Martins, Philippe},
  booktitle={2024 IEEE 100th Vehicular Technology Conference (VTC2024-Fall)}, 
  title={Optimizing LGBM for Multi-Classification of 5G SA Traffic}, 
  year={2024},
  volume={},
  number={},
  pages={1-7},
  doi={10.1109/VTC2024-Fall63153.2024.10757717}}

@misc{yoon2024scalablerobustmobileactivity,
      title={Scalable and Robust Mobile Activity Fingerprinting via Over-the-Air Control Channel in 5G Networks}, 
      author={Gunwoo Yoon and Byeongdo Hong},
      year={2024},
      eprint={2409.12572},
      archivePrefix={arXiv},
      primaryClass={cs.NI}
}

@ARTICLE{9003304,
  author={Meneghello, Francesca and Rossi, Michele and Bui, Nicola},
  journal={IEEE Network}, 
  title={Smartphone Identification via Passive Traffic Fingerprinting: A Sequence-to-Sequence Learning Approach}, 
  year={2020},
  volume={34},
  number={2},
  pages={112-120},
  doi={10.1109/MNET.001.1900101}}

@INPROCEEDINGS{9500470,
  author={Pelati, Annalisa and Meo, Michela and Dini, Paolo},
  booktitle={ICC 2021 - IEEE International Conference on Communications}, 
  title={A Semi-supervised Method to Identify Urban Anomalies through LTE PDCCH Fingerprinting}, 
  year={2021},
  volume={},
  number={},
  pages={1-6},
  doi={10.1109/ICC42927.2021.9500470}}

@inproceedings{10.1145/3448300.3467826,
author = {Chlosta, Merlin and Rupprecht, David and P\"{o}pper, Christina and Holz, Thorsten},
title = {5G SUCI-catchers: still catching them all?},
year = {2021},
isbn = {9781450383493},
publisher = {Association for Computing Machinery},
address = {New York, NY, USA},
doi = {10.1145/3448300.3467826},
booktitle = {Proceedings of the 14th ACM Conference on Security and Privacy in Wireless and Mobile Networks},
pages = {359–364},
numpages = {6},
location = {Abu Dhabi, United Arab Emirates},
series = {WiSec '21}
}

@inproceedings{DBLP:conf/ndss/HussainECLB19,
  author       = {Syed Rafiul Hussain and
                  Mitziu Echeverria and
                  Omar Chowdhury and
                  Ninghui Li and
                  Elisa Bertino},
  title        = {Privacy Attacks to the 4G and 5G Cellular Paging Protocols Using Side
                  Channel Information},
  booktitle    = {26th Annual Network and Distributed System Security Symposium, {NDSS}
                  2019, San Diego, California, USA, February 24-27, 2019},
  publisher    = {The Internet Society},
  year         = {2019}
}

@inproceedings{hong2026passive,
  title={Passive Multi-Target {GUTI} Identification via Visual-{RF} Correlation in {LTE} Networks},
  author={Hong, Byeongdo and Yoon, Gunwoo},
  booktitle={Proceedings of the Network and Distributed System Security Symposium (NDSS)},
  year={2026},
  month={March},
  address={San Diego, CA, USA},
  publisher={Internet Society}}

@inproceedings{10.5555/3766078.3766355,
author = {Luo, Shijie and Garbelini, Matheus E. and Chattopadhyay, Sudipta and Zhou, Jianying},
title = {SNI5GECT: a practical approach to inject aNRchy into 5G NR},
year = {2025},
isbn = {978-1-939133-52-6},
publisher = {USENIX Association},
address = {USA},
booktitle = {Proceedings of the 34th USENIX Conference on Security Symposium},
articleno = {277},
numpages = {20},
location = {Seattle, WA, USA},
series = {SEC '25}
}

@inproceedings{10.1145/3680121.3697808,
author = {Wan, Haoran and Cao, Xuyang and Marder, Alexander and Jamieson, Kyle},
title = {NR-Scope: A Practical 5G Standalone Telemetry Tool},
year = {2024},
isbn = {9798400711084},
publisher = {Association for Computing Machinery},
address = {New York, NY, USA},
doi = {10.1145/3680121.3697808},
booktitle = {Proceedings of the 20th International Conference on Emerging Networking EXperiments and Technologies},
pages = {73–80},
numpages = {8},
location = {Los Angeles, CA, USA},
series = {CoNEXT '24}
}

@inproceedings{DBLP:conf/uss/NiL0ZX23,
  author       = {Tao Ni and
                  Guohao Lan and
                  Jia Wang and
                  Qingchuan Zhao and
                  Weitao Xu},
  editor       = {Joseph A. Calandrino and
                  Carmela Troncoso},
  title        = {Eavesdropping Mobile App Activity via Radio-Frequency Energy Harvesting},
  booktitle    = {32nd {USENIX} Security Symposium, {USENIX} Security 2023, Anaheim,
                  CA, USA, August 9-11, 2023},
  pages        = {3511--3528},
  publisher    = {{USENIX} Association},
  year         = {2023}
}

@inproceedings{DBLP:conf/ndss/0002CH0H25,
  author       = {Rui Xiao and
                  Xiankai Chen and
                  Yinghui He and
                  Jun Han and
                  Jinsong Han},
  title        = {Lend Me Your Beam: Privacy Implications of Plaintext Beamforming Feedback
                  in WiFi},
  booktitle    = {32nd Annual Network and Distributed System Security Symposium, {NDSS}
                  2025, San Diego, California, USA, February 24-28, 2025},
  publisher    = {The Internet Society},
  year         = {2025}
}

@inproceedings{DBLP:conf/uss/JedermannSLS24,
  author       = {Eric Jedermann and
                  Martin Strohmeier and
                  Vincent Lenders and
                  Jens B. Schmitt},
  editor       = {Davide Balzarotti and
                  Wenyuan Xu},
  title        = {{RECORD:} {A} RECeption-Only Region Determination Attack on {LEO}
                  Satellite Users},
  booktitle    = {33rd {USENIX} Security Symposium, {USENIX} Security 2024, Philadelphia,
                  PA, USA, August 14-16, 2024},
  publisher    = {{USENIX} Association},
  year         = {2024}
}

@inproceedings{DBLP:conf/uss/KotuliakELRC22,
  author       = {Martin Kotuliak and
                  Simon Erni and
                  Patrick Leu and
                  Marc R{\"{o}}schlin and
                  Srdjan Capkun},
  editor       = {Kevin R. B. Butler and
                  Kurt Thomas},
  title        = {LTrack: Stealthy Tracking of Mobile Phones in {LTE}},
  booktitle    = {31st {USENIX} Security Symposium, {USENIX} Security 2022, Boston,
                  MA, USA, August 10-12, 2022},
  pages        = {1291--1306},
  publisher    = {{USENIX} Association},
  year         = {2022}
}

@inproceedings{DBLP:conf/uss/LakshmananBKCH21,
  author       = {Nitya Lakshmanan and
                  Nishant Budhdev and
                  Min Suk Kang and
                  Mun Choon Chan and
                  Jun Han},
  editor       = {Michael D. Bailey and
                  Rachel Greenstadt},
  title        = {A Stealthy Location Identification Attack Exploiting Carrier Aggregation
                  in Cellular Networks},
  booktitle    = {30th {USENIX} Security Symposium, {USENIX} Security 2021, August 11-13,
                  2021},
  pages        = {3899--3916},
  publisher    = {{USENIX} Association},
  year         = {2021}
}

@article{DBLP:journals/corr/abs-2512-20622,
  author       = {Atmane Ayoub Mansour Bahar and
                  Andr{\'{e}}s Alay{\'{o}}n Glazunov and
                  Romaric Duvignau},
  title        = {How Feasible are Passive Network Attacks on 5G Networks and Beyond?
                  {A} Survey},
  journal      = {CoRR},
  volume       = {abs/2512.20622},
  year         = {2025},
  doi          = {10.48550/ARXIV.2512.20622},
  eprinttype   = {arXiv},
  eprint       = {2512.20622}
}

@inproceedings{jawne2025aiassisted,
  author    = {Aishwarya Jawne and
               Georgios Sklivanitis and
               Dimitris A. Pados and
               Elizabeth Serena Bentley},
  title     = {AI-Assisted RF Fingerprinting for Identification of User Devices in 5G and FutureG},
  booktitle = {Workshop on Security and Privacy of Next-Generation Networks (FutureG) 2025},
  year      = {2025},
  month     = {February},
  address   = {San Diego, CA, USA},
  doi       = {10.14722/futureg.2025.23009},
  note      = {Presented at NDSS Symposium Workshop}
}

@inproceedings{hoang:ltesniffer,
  title = {{LTESniffer: An Open-source LTE Downlink/Uplink Eavesdropper}},
  author = {Hoang, Dinh Tuan and Park, CheolJun and Son, Mincheol and Oh, Taekkyung and Bae, Sangwook and Ahn, Junho and Oh, BeomSeok and Kim, Yongdae},
  booktitle = {16th ACM Conference on Security and Privacy in Wireless and Mobile Networks (WiSec '23)},
  year = {2023}
}

@inproceedings {236354,
author = {Hojoon Yang and Sangwook Bae and Mincheol Son and Hongil Kim and Song Min Kim and Yongdae Kim},
title = {Hiding in Plain Signal: Physical Signal Overshadowing Attack on {LTE}},
booktitle = {28th USENIX Security Symposium (USENIX Security 19)},
year = {2019},
isbn = {978-1-939133-06-9},
address = {Santa Clara, CA},
pages = {55--72},
publisher = {USENIX Association},
month = aug
}

@inproceedings{10.1145/3495243.3560525,
author = {Erni, Simon and Kotuliak, Martin and Leu, Patrick and Roeschlin, Marc and Capkun, Srdjan},
title = {AdaptOver: adaptive overshadowing attacks in cellular networks},
year = {2022},
isbn = {9781450391818},
publisher = {Association for Computing Machinery},
address = {New York, NY, USA},
doi = {10.1145/3495243.3560525},
booktitle = {Proceedings of the 28th Annual International Conference on Mobile Computing And Networking},
pages = {743–755},
numpages = {13},
location = {Sydney, NSW, Australia},
series = {MobiCom '22}
}

@INPROCEEDINGS{11094472,
  author={Mi, Ze-Yu and Yang, Yu-Bin},
  booktitle={2025 IEEE/CVF Conference on Computer Vision and Pattern Recognition (CVPR)}, 
  title={ADD: Attribution-Driven Data Augmentation Framework for Boosting Image Super-Resolution}, 
  year={2025},
  volume={},
  number={},
  pages={23101-23110},
  doi={10.1109/CVPR52734.2025.02151}}

@inproceedings{DBLP:conf/aaai/LuXWDH26,
  author       = {Zhiguang Lu and
                  Qianqian Xu and
                  Peisong Wen and
                  Siran Dai and
                  Qingming Huang},
  editor       = {Sven Koenig and
                  Chad Jenkins and
                  Matthew E. Taylor},
  title        = {HiGFA: Hierarchical Guidance for Fine-grained Data Augmentation with
                  Diffusion Models},
  booktitle    = {Fortieth {AAAI} Conference on Artificial Intelligence, Thirty-Eighth
                  Conference on Innovative Applications of Artificial Intelligence,
                  Sixteenth Symposium on Educational Advances in Artificial Intelligence,
                  {AAAI} 2026, Singapore, January 20-27, 2026},
  pages        = {7600--7608},
  publisher    = {{AAAI} Press},
  year         = {2026}
}

@techreport{3gpp.38.214,
  author       = {3GPP},
  title        = {{3GPP TS 38.214}: NR; Physical layer procedures for data},
  institution  = {3rd Generation Partnership Project (3GPP)},
  year         = {2024},
  note         = {Version dependent on Release (e.g., V18.x for Rel-18)},
  howpublished = {3GPP Technical Specification}
}

@article{DBLP:journals/ejwcn/HaghrahAAN23,
  author       = {Amiraslan Haghrah and
                  Mehran Pourmohammad Abdollahi and
                  Hosein Azarhava and
                  Javad Musevi Niya},
  title        = {A survey on the handover management in 5G-NR cellular networks: aspects,
                  approaches and challenges},
  journal      = {{EURASIP} J. Wirel. Commun. Netw.},
  volume       = {2023},
  number       = {1},
  pages        = {52},
  year         = {2023}
}

@INPROCEEDINGS{7510709,
  author={Arshad, Rabe and ElSawy, Hesham and Sorour, Sameh and Al-Naffouri, Tareq Y. and Alouini, Mohamed-Slim},
  booktitle={2016 IEEE International Conference on Communications (ICC)}, 
  title={Handover management in dense cellular networks: A stochastic geometry approach}, 
  year={2016},
  volume={},
  number={},
  pages={1-7},
  doi={10.1109/ICC.2016.7510709}}

@article{DBLP:journals/tifs/YangL24,
  author       = {Xuan Yang and
                  Dongming Li},
  title        = {{LED-RFF:} {LTE} DMRS-Based Channel Robust Radio Frequency Fingerprint
                  Identification Scheme},
  journal      = {{IEEE} Trans. Inf. Forensics Secur.},
  volume       = {19},
  pages        = {1855--1869},
  year         = {2024},
  doi          = {10.1109/TIFS.2023.3343079}
}

@article{DBLP:journals/corr/abs-2512-10809,
  author       = {Reinhard Wiesmayr and
                  Frederik Zumegen and
                  Sueda Taner and
                  Chris Dick and
                  Christoph Studer},
  title        = {CSI-Based User Positioning, Channel Charting, and Device Classification
                  with an {NVIDIA} 5G Testbed},
  journal      = {CoRR},
  volume       = {abs/2512.10809},
  year         = {2025},
  doi          = {10.48550/ARXIV.2512.10809},
  eprinttype   = {arXiv},
  eprint       = {2512.10809}
}

@misc{srsRAN,
  title        = {srsRAN},
  author       = {Software Radio Systems},
  howpublished = {\url{https://github.com/jsroldan/srsRAN}},
  note         = {Accessed: 2026-04-25}
}

@misc{open5gs,
  title        = {open5gs},
  author       = {open5gs},
  howpublished = {\url{https://github.com/open5gs/open5gs}},
  note         = {Accessed: 2026-04-25}
}

@article{DBLP:journals/iotj/PengPFL24,
  author       = {Linning Peng and
                  Haichuan Peng and
                  Hua Fu and
                  Ming Liu},
  title        = {Channel-Robust Radio Frequency Fingerprint Identification for Cellular
                  Uplink {LTE} Devices},
  journal      = {{IEEE} Internet Things J.},
  volume       = {11},
  number       = {10},
  pages        = {17154--17169},
  year         = {2024},
  doi          = {10.1109/JIOT.2024.3358904}
}

@article{DBLP:journals/wcl/JangKKMKK25,
  author       = {Daegun Jang and
                  Gayeon Kim and
                  Byeong{-}Gwon Kang and
                  Kyungsik Min and
                  Youngju Kim and
                  Taehyoung Kim},
  title        = {Convolutional Neural Network-Based {PUSCH} {DMRS} Pattern Optimization
                  for 5G New Radio},
  journal      = {{IEEE} Wirel. Commun. Lett.},
  volume       = {14},
  number       = {10},
  pages        = {3299--3303},
  year         = {2025},
  doi          = {10.1109/LWC.2025.3592297}
}

@inproceedings{DBLP:conf/iclr/DufumierNTT25,
  author       = {Benoit Dufumier and
                  Javiera Castillo Navarro and
                  Devis Tuia and
                  Jean{-}Philippe Thiran},
  title        = {What to align in multimodal contrastive learning?},
  booktitle    = {The Thirteenth International Conference on Learning Representations,
                  {ICLR} 2025, Singapore, April 24-28, 2025},
  publisher    = {OpenReview.net},
  year         = {2025}
}

@inproceedings{DBLP:conf/cvpr/Duan0PLS025,
  author       = {Siyuan Duan and
                  Yuan Sun and
                  Dezhong Peng and
                  Zheng Liu and
                  Xiaomin Song and
                  Peng Hu},
  title        = {Fuzzy Multimodal Learning for Trusted Cross-modal Retrieval},
  booktitle    = {{IEEE/CVF} Conference on Computer Vision and Pattern Recognition,
                  {CVPR} 2025, Nashville, TN, USA, June 11-15, 2025},
  pages        = {20747--20756},
  publisher    = {Computer Vision Foundation / {IEEE}},
  year         = {2025}
}

@misc{SeleniumHQ2026,
  author       = {{SeleniumHQ}},
  title        = {Selenium},
  howpublished = {\url{https://github.com/SeleniumHQ/selenium}},
  year         = {2026},
  note         = {A browser automation framework and ecosystem}
}

@misc{PCAPdroid2026,
  author       = {{emanuele-f}},
  title        = {{PCAPdroid}: No-root network monitor, firewall and PCAP dumper for Android},
  howpublished = {\url{https://github.com/emanuele-f/PCAPdroid}},
  year         = {2026},
  note         = {Accessed: June 2026}
}

@techreport{3gpp.38.213,
  author       = {3GPP},
  title        = {NR; Physical layer procedures for control},
  institution  = {3GPP},
  year         = {2024},
  type         = {Technical Specification},
  number       = {TS 38.213}
}

@inproceedings{DBLP:conf/ndss/ShaikSBAN16,
  author       = {Altaf Shaik and
                  Jean{-}Pierre Seifert and
                  Ravishankar Borgaonkar and
                  N. Asokan and
                  Valtteri Niemi},
  title        = {Practical Attacks Against Privacy and Availability in 4G/LTE Mobile
                  Communication Systems},
  booktitle    = {23rd Annual Network and Distributed System Security Symposium, {NDSS}
                  2016, San Diego, California, USA, February 21-24, 2016},
  publisher    = {The Internet Society},
  year         = {2016}
}

@inproceedings{DBLP:conf/ccs/SirinamIJW18,
  author       = {Payap Sirinam and
                  Mohsen Imani and
                  Marc Juarez and
                  Matthew Wright},
  editor       = {David Lie and
                  Mohammad Mannan and
                  Michael Backes and
                  XiaoFeng Wang},
  title        = {Deep Fingerprinting: Undermining Website Fingerprinting Defenses with
                  Deep Learning},
  booktitle    = {Proceedings of the 2018 {ACM} {SIGSAC} Conference on Computer and
                  Communications Security, {CCS} 2018, Toronto, ON, Canada, October
                  15-19, 2018},
  pages        = {1928--1943},
  publisher    = {{ACM}},
  year         = {2018}
}

@inproceedings {287133,
author = {Renjie Xie and Jiahao Cao and Enhuan Dong and Mingwei Xu and Kun Sun and Qi Li and Licheng Shen and Menghao Zhang},
title = {Rosetta: Enabling Robust {TLS} Encrypted Traffic Classification in Diverse Network Environments with {TCP-Aware} Traffic Augmentation},
booktitle = {32nd USENIX Security Symposium (USENIX Security 23)},
year = {2023},
isbn = {978-1-939133-37-3},
address = {Anaheim, CA},
pages = {625--642},
publisher = {USENIX Association},
month = aug
}

@article{DBLP:journals/popets/RahmanSMG020,
  author       = {Mohammad Saidur Rahman and
                  Payap Sirinam and
                  Nate Mathews and
                  Kantha Girish Gangadhara and
                  Matthew Wright},
  title        = {Tik-Tok: The Utility of Packet Timing in Website Fingerprinting Attacks},
  journal      = {Proc. Priv. Enhancing Technol.},
  volume       = {2020},
  number       = {3},
  pages        = {5--24},
  year         = {2020}
}

@inproceedings{DBLP:conf/ccs/SirinamMR019,
  author       = {Payap Sirinam and
                  Nate Mathews and
                  Mohammad Saidur Rahman and
                  Matthew Wright},
  editor       = {Lorenzo Cavallaro and
                  Johannes Kinder and
                  XiaoFeng Wang and
                  Jonathan Katz},
  title        = {Triplet Fingerprinting: More Practical and Portable Website Fingerprinting
                  with N-shot Learning},
  booktitle    = {Proceedings of the 2019 {ACM} {SIGSAC} Conference on Computer and
                  Communications Security, {CCS} 2019, London, UK, November 11-15, 2019},
  pages        = {1131--1148},
  publisher    = {{ACM}},
  year         = {2019}
}

@ARTICLE{8371237,
  author={Ahmed, Irfan and Khammari, Hedi and Shahid, Adnan and Musa, Ahmed and Kim, Kwang Soon and De Poorter, Eli and Moerman, Ingrid},
  journal={IEEE Communications Surveys \& Tutorials}, 
  title={A Survey on Hybrid Beamforming Techniques in 5G: Architecture and System Model Perspectives}, 
  year={2018},
  volume={20},
  number={4},
  pages={3060-3097},
  doi={10.1109/COMST.2018.2843719}
}

@ARTICLE{9034044,
  author={Sim, Min Soo and Lim, Yeon-Geun and Park, Sang Hyun and Dai, Linglong and Chae, Chan-Byoung},
  journal={IEEE Access}, 
  title={Deep Learning-Based mmWave Beam Selection for 5G NR/6G With Sub-6 GHz Channel Information: Algorithms and Prototype Validation}, 
  year={2020},
  volume={8},
  number={},
  pages={51634-51646},
  doi={10.1109/ACCESS.2020.2980285}
}

@techreport{3GPP_TS_38_213,
  author       = {{3GPP}},
  title        = {{TS 38.213: NR; Physical layer procedures for control}},
  institution  = {3rd Generation Partnership Project (3GPP)},
  year         = {2026},
  number       = {38.213}
}

@techreport{3GPP_TS_38_212,
  author       = {{3GPP}},
  title        = {{TS 38.212: NR; Multiplexing and channel coding}},
  institution  = {3rd Generation Partnership Project (3GPP)},
  year         = {2026},
  number       = {38.212}
}





\end{document}